\documentstyle[aps,prl,multicol,eqsecnum,psfig]{revtex}
\renewcommand{\narrowtext}{\begin{multicols}{2} \global\columnwidth20.5pc}
\renewcommand{\widetext}{\end{multicols} \global\columnwidth42.5pc} 

\def\top#1{\vskip #1\begin{picture}(290,80)(80,500)\thinlines \put(
65,500){\line( 1, 0){255}}\put(320,500){\line( 0, 1){
5}}\end{picture}}
\def\bottom#1{\vskip #1\begin{picture}(290,80)(80,500)\thinlines \put(
330,500){\line( 1, 0){255}}\put(330,500){\line( 0, -1){
5}}\end{picture}}

\begin{document}

\title{Exact Zero Temperature Correlation Functions
for Two-Leg Hubbard Ladders and Carbon Nanotubes}
\author{Robert Konik and Andreas W.W. Ludwig}
\address{Department of Physics, UCSB, Santa Barbara CA 93106-4030}
\date{\today}
\maketitle

\begin{abstract}
Motivated by recent work of Lin, Balents, and Fisher \cite{lin},
we compute correlation functions at zero temperature for 
weakly coupled 
two-leg Hubbard ladders and $(N,N)$ armchair carbon nanotubes.
In \cite{lin} it was argued that such systems
renormalize towards the $SO(8)$ Gross-Neveu model, an integrable theory.
We exploit this integrability to perform the computation at the $SO(8)$
invariant point.  Any terms breaking the $SO(8)$ symmetry can be treated
systematically in perturbation theory, leading
to a model with same qualitative features as the integrable theory.
Using said correlators,
we determine the optical
conductivity, the single-particle spectral function, 
and the I-V curve for
tunneling into the system from an external metallic lead.  The 
frequency, $\omega$,
dependent optical conductivity
is determined exactly for  $\omega < 3m$
($m$ being the fermion particle mass in the $SO(8)$ Gross-Neveu model).
It is characterized by a sharp ``exciton'' peak at $\omega = \sqrt{3}m$ ,
followed by the onset of the particle-hole continuum
beginning at $\omega = 2m$.  Interactions modify this onset to
$\sigma (\omega + 2m) \sim \omega^{1/2}$ 
and not the $\omega^{-1/2}$ one would expect
from the van-Hove singularity in the density of states.
Similarly, we obtain the exact single particle spectral function
for energies less than $3m$.
The latter possesses
a delta function peak arising from single particle
excitations, together with a two-particle continuum for $\omega\geq 2m$.
The final quantity we compute is
the tunneling I-V curve to lowest non-vanishing order in the tunneling
matrix elements.
For this quantity, we present exact results
for voltages, $V < (1+\sqrt{3})m$.
The resulting
differential conductance is marked by a finite
jump at $\omega = 2m$, the energy of the onset of tunneling into
the continuum of two particle states.  Through integrability,
we are able to characterize this jump exactly. 

All calculations are done through
form-factor expansions of correlation functions.  These
give exact closed form expressions for spectral functions
because the $SO(8)$ Gross-Neveu model
is massive: each term in the expansion has an
energy threshold below which it does not contribute. Thus, we obtain exact 
results below certain thresholds by computing a finite
number of terms in this series.
Previous to this paper, the only computed form-factor of $SO(8)$ 
Gross-Neveu
was the two particle form factor of an $SO(8)$ current 
with two fundamental
fermions.  In this paper we compute the set of all one and two
particle form factors for all relevant fields, the currents as well as
the kinks and fermions.
\end{abstract}
\pacs{PACS numbers: ????}

%
%
%
\newcommand{\lb}{\langle}
\newcommand{\rb}{\rangle}
%
%
\newcommand{\zb}{\bar{z} }
\newcommand{\half}{{1 \over 2}}
%
%
\newcommand{\th}{\theta}         \newcommand{\Th}{\Theta}
\newcommand{\ga}{\gamma}         \newcommand{\Ga}{\Gamma}
\newcommand{\be}{\beta}
\newcommand{\al}{\alpha}
\newcommand{\ep}{\varepsilon}
\newcommand{\la}{\lambda}        \newcommand{\La}{\Lambda}
\newcommand{\de}{\delta}         \newcommand{\De}{\Delta}
\newcommand{\om}{\omega}         \newcommand{\Om}{\Omega}
\newcommand{\sig}{\sigma}        \newcommand{\Sig}{\Sigma}
\newcommand{\vphi}{\varphi}
%
%
\newcommand{\CA}{{\cal A}}       
\newcommand{\CB}{{\cal B}}       
\newcommand{\CC}{{\cal C}}
\newcommand{\CD}{{\cal D}}       
\newcommand{\CE}{{\cal E}}       
\newcommand{\CF}{{\cal F}}
\newcommand{\CG}{{\cal G}}       
\newcommand{\CH}{{\cal H}}       
\newcommand{\CI}{{\cal J}}
\newcommand{\CJ}{{\cal J}}       
\newcommand{\CK}{{\cal K}}       
\newcommand{\CL}{{\cal L}}
\newcommand{\CM}{{\cal M}}       
\newcommand{\CN}{{\cal N}}       
\newcommand{\CO}{{\cal O}}
\newcommand{\CP}{{\cal P}}       
\newcommand{\CQ}{{\cal Q}}       
\newcommand{\CR}{{\cal R}}
\newcommand{\CS}{{\cal S}}       
\newcommand{\CT}{{\cal T}}       
\newcommand{\CU}{{\cal U}}
\newcommand{\CV}{{\cal V}}       
\newcommand{\CW}{{\cal W}}       
\newcommand{\CX}{{\cal X}}
\newcommand{\CY}{{\cal Y}}       
\newcommand{\CZ}{{\cal Z}}

\newcommand{\rvac}{\hbox{$\vert 0\rangle$}}
\newcommand{\lvac}{\hbox{$\langle 0 \vert $}}

%
%
%
\font\numbers=cmss12
\font\upright=cmu10 scaled\magstep1
\newcommand{\stroke}{\vrule height8pt width0.4pt depth-0.1pt}
\newcommand{\topfleck}{\vrule height8pt width0.5pt depth-5.9pt}
\newcommand{\botfleck}{\vrule height2pt width0.5pt depth0.1pt}
\newcommand{\Z}{\ifmmode\Zmath\else$\Zmath$\fi}
\newcommand{\Q}{\ifmmode\Qmath\else$\Qmath$\fi}
\newcommand{\N}{\ifmmode\Nmath\else$\Nmath$\fi}
\newcommand{\C}{\ifmmode\Cmath\else$\Cmath$\fi}
\newcommand{\R}{\ifmmode\Rmath\else$\Rmath$\fi}


\newcommand{\del}{\partial}
\newcommand{\js}{J_s}
\newcommand{\rs}{R_s}
\newcommand{\bjs}{\bar{J}_s}
\newcommand{\brs}{\bar{R}_s}
\newcommand{\jnm}{J^{n,m}_s}
\newcommand{\rnm}{R^{n,m}_s}

\newcommand{\prnm}{R^{n',m'}_s}
\newcommand{\bjnm}{\bar{J}^{n,m}_s}
\newcommand{\brnm}{\bar{R}^{n,m}_s}
\newcommand{\pbrnm}{\bar{R}^{n',m'}_s}
\newcommand{\bt}{\beta}
\newcommand{\bs}{{\hat{\beta}^2}}
\newcommand{\tcor}{\langle T(G^{12}_1(x,\tau)G^{12}_1(0,0))\rangle}
\newcommand{\cor}{\langle G^{12}_1(x,t)G^{12}_1(0,0)\rangle}

\narrowtext 
\section{Introduction}

Strongly interacting systems are a focal point of modern solid state
physics.  The behaviour of these systems is often not captured by 
approximations, 
based on various `free particle' non-interacting models,
since in the presence of generically strong
interactions the physics is typically much richer, exhibiting
qualitatively new features.
Calculational tools to 
analyze the behaviour of strongly interacting theories are
few in number.  Amongst these are the tools of
integrability, which are available for
a number of special $1+1$-dimensional models.
These methods allow for exact and detailed computations
of many physically relevant properties of these theories.
But as integrable models have to be of a special form,
i.e. need to be `fine-tuned',
it is often believed that these tools are of little relevance
for systems observed in the laboratory, since the latter would never
generically be of the particular form necessary for
integrability.

However this need not be the case.  Recently,
in the context of transport experiments
on tunneling point contacts in fractional quantum Hall effect 
devices \cite{Milliken}, 
it has been pointed out by \cite{FLS} 
that when renormalization group (RG) thinking is combined 
with integrability,
a special and `fine-tuned' integrable Hamiltonian may actually describe
the behaviour of a realistic system.
The Hamiltonian of a generic system may be attracted under RG
transformations to an integrable one, which then can be
analyzed using the methods of integrability.
In this way these powerful methods can be brought to
bear on experimentally realizable systems and phenomena,
and their predictions be directly compared with experimental
data.  

In this paper we address another set of examples of a similar spirit.
These are quasi one-dimensional interacting electronic systems,
of which two prominent experimental realizations
are two-leg Hubbard ladders (see for example \cite{Dagotto})
and single-walled carbon
nanotubes \cite{Ebbesen}.

Two-leg ladders have recently been the focus of much
theoretical and experimental activity.  At half-filling
they are Mott insulators, exhibiting gaps to all excitations,
and in particular a spin gap.  These are typical
examples of `spin-liquids'.  Upon doping, the gaps
to all excitations
except for those with charge-two survive \cite{doped}.
The gapless charge modes induce
quasi long-range
superconducting pairing correlations, with approximate
$d$-wave symmetry, reminiscent of underdoped cuprate superconductors.

Carbon nanotubes are novel materials whose mechanical and
electronic properties promise potential for new technological
applications \cite{Ebbesen}.
They are formed by wrapping graphite sheets into cylinders of nanoscale
dimensions.  They support electronic excitations, which,
for a prominent member of the nanotube family, the armchair $(n,n)$ type,
can be described by the same theoretical model as that
used for the two-leg Hubbard ladders \cite{bal,kro}.  Even though
these systems would be one-dimensional band metals
in the absence of interactions,
they become Mott insulators at half-filling due to the presence of
short-ranged electronic interactions, which play an important role due
to their one-dimensional nature.  It is these interaction
effects that we analyze exactly in this paper using the
powerful methods of integrability.
After experimental techniques
had been developed to fabricate long single-walled
nanotubes with high yields in the laboratory, this field
of material science has seen a explosive
development \cite{nanorefs}.
Electronic
properties can be measured relatively easily by attaching
metallic leads \cite{twoterminal}
or by  tunneling into these materials with 
scanning tunneling microscope (STM)
tips.  The practical feasibility of such tunneling
experiments from an STM tip into an individual single-walled
nanotube placed upon a gold substrate (screening long-range
Coulomb forces) has recently been demonstrated by J.W.G. Wild\"oer et al.
\cite{nanorefs}.
Thus, the predictions that we make in this paper,
especially those for the tunneling $I(V)$-curve, have
direct bearing on possible future experiments on
these materials.

As said, both systems, two-leg ladders and single-walled armchair nanotubes,
would be one-dimensional metals in the absence of electron
interactions. These
are described theoretically (on scales
much smaller than the non-interacting band width) by two species
of spinful massless
Dirac fermions in $(1+1)$ dimensions\footnote{These two species
have equal Fermi velocities due to particle-hole symmetry
present at half-filling.}.  For the ladder compounds,
the two species arise in an obvious way from the two rungs of
the ladder, whereas for the nanotubes they arise from the
particular band structure of the underlying hexagonal
graphite lattice, characterized by two Fermi points in the Brillouin
zone \cite{bal,kro}.  These massless  Fermi surface excitations
interact with short-range interactions whose
detailed nature is determined by non-universal
microscopic considerations.

A notable observation was made recently by Lin, Balents, 
and Fisher \cite{lin}.
These authors argued
that within an 1-loop RG any such model with generic, non-chiral,
short range interactions
flows at half-filling into a theory
with an immense symmetry, namely the $SO(8)$ symmetric Gross-Neveu
model.  This model not only has a large 
$SO(8)$ global symmetry, which encompasses an one-dimensional
version of $SO(5)$ recently advocated by 
S. C. Zhang~\cite{ZhangSO5},
but in addition has an infinite number of hidden
conservation laws, which are a consequence of the
integrability of this model.
The $SO(5)$ subgroup symmetry is the same as
that studied by \cite{ScalapinoZhang} in the context of lattice models of
interacting two-leg ladders.  We point out that it would be
straightforward to construct a corresponding $SO(8)$ invariant lattice
model of a two-leg ladder.

The use of the RG in \cite{lin} can be understood in the following sense.
Since the analysis in \cite{lin} is based on an 1-loop RG,
the initial microscopic (bare)
interactions must be small enough so that
the integrable $SO(8)$ invariant RG trajectory is approached sufficiently
closely after a number of RG steps, before leaving the
range of validity of the 1-loop RG equations.
Whenever this is the case, it is argued that 
the integrable model is approached
independently of the (sufficiently weak) values of the bare interactions.
The situation for the 2-leg ladder
is thus similar in spirit
to that of the point contact device encountered
in \cite{FLS},  where only a single operator was relevant, and
this relevant operator
was integrable. In the latter case all other interactions were irrelevant
in the RG sense, and could in principle be treated perturbatively.

The requirement of the 
RG that the interactions be short-ranged is natural
in the case of the Hubbard ladders.  However it may not seem so
in the case of the carbon nanotubes.  Recent theory 
\cite{lutthe} and experiment \cite{lutexp}
have discussed the case where long-ranged Coulomb forces drive Luttinger
liquid behaviour in single-walled carbon nanotubes.  
However we do not have such
situations in mind for the paper at hand.  Rather we want to consider
situations such as those found in the experiments of
J.W.G. Wild\"oer et al. \cite{nanorefs} where the long range forces
are screened.

Although the restriction to such experiments in the case of the carbon
nanotubes
places us upon safe ground, it is not 
inconceivable that experiments where
the long-ranged forces are present would nevertheless see 
behaviour indicative of the SO(8) symmetry.  
An unscreened force translates into an unusually
large bare coupling (in comparison with other bare couplings)
in the forward scattering direction.
However this does not mean the RG is inapplicable.  The RG still 
indicates a potential
enhancement in the symmetry.  Because of the large
bare coupling, the RG must be  
run a longer time before any enhancement
would be seen 
but nevertheless an enhancement may well occur at some
low energy scale.  In terms of the experiments in \cite{lutexp}, this would
mean that at medium energy scales, 
Luttinger liquid behaviour would predominate,
while at much lower energy scales, SO(8) behaviour would be expected.
However at current standing, the material science is not advanced
to the point where it is possible to 
accurately probe the very low energy behaviour.
But the potential for advancement in this area is ever present.

The RG analysis further requires the bare couplings to be weak.  With
Hubbard ladder compounds, this condition will not be generically met,
although it certainly will not be universally violated.  However with
(N,N)-armchair carbon nanotubes, the bare couplings are naturally 
weak.  It is one of the hallmarks of the physics of the (N,N) armchair
carbon nanotubes that the electrons are delocalized around the circumference
of the tube.  This in turn leads to a scaling of the effective short-ranged
interaction by 1/N, making it naturally small \cite{bal}.

It can, however, be questioned on a more fundamental level
whether an 1-loop RG adequately describes
the system's behaviour.  Difficulties with the analysis in \cite{lin} 
take two forms.
As a first objection,
the authors of \cite{tsv} point out that an RG flow can imply
a symmetry restoration which in fact does not occur.  As an example
they consider
a U(1) symmetric Thirring model,
\begin{equation}\label{eIi}
{\cal{L}} = 
\bar\Psi_\alpha \gamma^\mu\partial_\mu\Psi_\alpha + {1\over 4}g_\parallel
(j_z)^2 + {1\over 4}g_\perp [(j_x)^2+(j_y)^2],
\end{equation}
where $j^\mu_a = \bar\Psi\gamma^\mu\sigma_a\Psi$.
Although the 1-loop RG equations for this model 
seems to indicate a generic
symmetry 
restoration to a more symmetric SU(2) case (i.e. $g_\perp = g_\parallel$), 
this in fact only occurs in a certain region of coupling space.
For $\pi - |g_\perp| > -g_\parallel > |g_\perp| > 0$, 
the U(1) model maps onto the sine-Gordon model with interaction 
$\cos (\beta\phi)$
\cite{Wieg}, where $\beta$ is given by
\begin{equation}\label{eIii}
\beta^2 = 8\pi - 8\mu ; ~~~~~ \mu = 
{\cos ^{-1} [\cos (g_\parallel ) / \cos (g_\perp)]}.
\end{equation}
The value of $\beta$ completely characterizes the model.  While
$g_\perp$ and $g_\parallel$ flow under the RG, the particular combination
of these parameters forming $\beta$ does not.
Thus for this particular region of parameter space
the model moves no closer to the SU(2) symmetric point
under an RG flow.  In other regions however (for example 
$|g_\perp| > |g_\parallel|$), 
the situation is better; the effect of the anisotropy in the 
couplings is exponentially suppressed.

However 
it is reasonably clear that such pessimism is not warranted in the analysis
of the RG of \cite{lin}.  A salient criticism of \cite{tsv} is that in
considering the action of the renormalization group, they fail
to consider the consequences of working in the scaling limit.  The
scaling limit is exactly the limit in which a field theory becomes
available.  In turn, the scaling limit places constraints upon the
possible range of bare couplings consistent with a field theory.
In the case of sine-Gordon, the underlying integrability/solvability
of the theory allows explicit investigation of this question.
It is found that the allowed range is such that even moderate
anisotropic deviations are forbidden \cite{anis}.  
The scaling limit, in other words,
enforces isotropy.  In cases where there are RG flows indicating an
enhancement in symmetry, this turns out to be a general phenomena and
it leads to an expanded notion of symmetry 
restoration \cite{anis}.

On a more concrete level, the breaking of the SU(2) symmetry considered in
\cite{tsv} is a rather special case.  
Ultimately, the parameter $\beta$ in the
sine-Gordon model is protected under an RG flow by the presence of a 
quantum group symmetry arrived at by deforming a Yangian symmetry present 
at the SU(2) point.  There is, however, no such known way to
deform the Yangian in $SO(8)$ Gross-Neveu.  Indeed the natural 
generalization 
of the sine-Gordon
model to $SO(8)$ is not to $SO(8)$ Gross-Neveu but to an affine toda 
$SO(8)$ theory where such a deformation of the Yangian symmetry is 
possible \cite{leclair}.

Another question that one must ask in looking at the analysis in \cite{tsv}
is how the choice of the symmetry breaking terms affects the symmetry
restoration.  The sine-Gordon model still possesses a U(1) symmetry.  
However
it is certainly possible to consider perturbations that break this U(1).
Such perturbations would destroy the quantum group symmetry of sine-Gordon
and thus might lead to symmetry restoration.  
This would be perhaps closer to the RG analysis
of \cite{lin} where a large number (nine) of marginal perturbations 
were included.  We in fact consider exactly such a situation in \cite{anis}
and find that indeed there is symmetry restoration.
\footnote{We note in passing that the authors of \cite{tsv} consider
an anisotropic Gross-Neveu model, a model of direct relevance
to the situation at hand.  They conclude
through a mean-field/large N limit 
computation that the model is intrinsically
anisotropic thus throwing doubts upon the analysis in Lin et al. \cite{lin}.
There appear to be 
several severe problems, however, with their
analysis.  The most telling is the manner in which the
bare couplings are scaled
in the large N limit.  The anisotropic model they consider
has three bare couplings.  One is chosen to not scale at all, one scales
as $1/N$, and the last scales as $1/N^{d/2}$, $d<1$.  With 
this scaling \cite{tsv}
the model is anisotropic.  However all that was 
done is to examine a system
with a diverging bare anisotropy and conclude that the effective theory
is similarly anisotropic.  And nowhere is the RG, which in this
model indicates symmetry restoration, allowed
to act.  As such, we believe this
example has little bearing on the situation at hand.}

The second objection to the analysis of \cite{lin} is 
its omission of chiral interactions
that alter the Fermi velocities \cite{kiv}.  
Such interactions, although they
are absent from the 1-loop RG, likely play a role 
at higher order.  However their effect is less drastic than envisioned
in \cite{kiv}.  There a scenario was considered where the
invariant RG trajectory of higher symmetry was inherently unstable to
perturbations.  However the $SO(8)$ RG ray in \cite{lin} has a basin of 
attraction of finite measure.  The effect of chiral interactions is to then
slightly alter the direction of the ray \cite{hsiu}.  
In turn the ratio of
masses of the various excitations are slightly perturbed away from one.

In taking account of these objections, prudence suggests
a modification in the understanding of the RG analysis of \cite{lin}.  
This analysis in \cite{lin} and \cite{hsiu} 
tells us that while the RG flow does
not restore an exact symmetry, it leaves us close to the symmetrical 
situation.  In particular, it indicates that while the masses in the actual
system may differ from their $SO(8)$ values, they do not wildly diverge.
One then understands the $SO(8)$ Gross-Neveu
theory, not as precisely
representative of the actual system, but in near perturbative vicinity
of it, that is,
as an excellent starting
point about which to perform perturbation theory in the
non-integrable interactions breaking $SO(8)$ in much
the same spirit as done for
a non-critical Ising model in the presence of a magnetic
field \cite{simo}.

In doing perturbation theory about $SO(8)$ Gross-Neveu it is
important to emphasize two salient points.  Firstly 
$SO(8)$ Gross-Neveu already contains all the basic features of the 
strongly interacting electron system.  
It thus provides a much better starting point for perturbation
theory than the two band metal from which it arose.  
In contrast, perturbation theory about the two band metal could not hope to
capture, even qualitatively, characteristics of the interacting system.
Secondly, $SO(8)$ Gross-Neveu is
a massive theory and so poses none of the attendant problems
of perturbation theory with a massless model (as a two band metal is).  
The perturbative series is well controlled;
the perturbed theory is connected in a continuous fashion 
to $SO(8)$ Gross-Neveu.

In this paper we focus on the integrable model at
half-filling and at zero temperature\footnote{The doped
system is examined in \cite{doped} .} 
and as such this work should be considered
as the starting point for the analysis of a perturbed $SO(8)$ Gross-Neveu
ladder/nanotube. (We do, however, in Section 3 
make a rough sketch of the effect
of perturbations on our results.  A more detailed analysis awaits further
work.)  Some consequences of the integrability
of the theory have
been exploited in \cite{lin}, notably those following
directly from  the structure of the
elementary excitations (`particles') of the theory.
If $U$ is the typical strength of
the interactions and $t$ a measure of the bandwidth
of the non-interacting model, the Gross-Neveu interactions
generate dynamically a single mass scale 
$m \propto te^{-t/U}$.  Due to the integrability, the
masses of all excitations
are universally related to $m$,
universal mass ratios being known exactly. This, together
with the knowledge of the quantum number assignment to these
excitations, has been exploited in \cite{lin} to make,
{\it inter alia}, a number of predictions for correlations functions,
largely of a qualitative nature.

Due to its integrability, an immense amount of information
is known about this strongly interacting system.
In this paper, we put this information to
use, to compute the exact functional form of spectral
functions, at zero temperature
and on scales below certain thresholds (to be discussed below).
To this end we use
the form factor approach described in
\cite{smirbook}.
In particular, we have obtained exact expressions for
the optical conductivity,
the single-electron spectral function (experimentally
accessible via photoemission experiments), as well
as the tunneling density of states (experimentally
observable for example by measurement of the differential
conductance for tunneling from a metallic lead into a
nanotube \cite{nanorefs}).

We emphasize that our results are exact on energy scales
below certain thresholds.
In order to appreciate the significance of this, it
is important to recall that even in an integrable
system, correlation functions cannot in general
be computed exactly.  However if a mass gap is present for
all excitations (such as for the Gross-Neveu model)
exact results for spectral functions
can be obtained below certain thresholds.
This  can be easily understood from the basic ideas
underlying the form factor approach to correlation functions in
integrable systems, which we now briefly review.
The key feature of an integrable system is the exact knowledge of a basis
of eigenstates of the fully interacting Hamiltonian.
At the root of integrability is
a well defined notion of ``particles'', or ``elementary
excitations'' in the fully interacting system.  These particles
scatter off each other only with two-body $S$-matrices,
that is, all particle production processes are absent
and particle number is conserved.  This is due to special
conservation laws which exist in an integrable model, preventing
the decay of these particles.  In this sense, an integrable system
is similar to a Fermi liquid.  
An additional feature is that new particles
can arise as bound states of already existing ones.
However the total number of different types of particles is
finite which makes the system analytically tractable.
For the $SO(8)$ Gross-Neveu model
these particles consist of fermionic particle states of mass $m$,
namely one octet of fundamental fermions 
and two octets of kinks. 
The excitations of the ladder or nanotube, possessing the quantum
numbers of the electron,
are represented
by eight of these kinks. In addition there are
29 bosonic particle states organized as a rank-2 $SO(8)$ 
anti-symmetric tensor
and a singlet, all
of mass $\sqrt{3} m$.  These
are bound states of the fermionic mass-$m$ particles.
The exact eigenstates of the interacting Hamiltonian are
simply $n$-particle states,  $|n;s_n\rangle$, where
$s_n$ collectively denotes the momenta and species labels
of the $n$ particles\footnote{This
can be made more explicit by writing, $|n;s_n\rangle =$
$|(p_1, a_1); (p_2, a_2);...;  (p_n, a_n)\rangle $.}.

The form-factor representation of any correlation function
is obtained by inserting a resolution of the identity
corresponding to the basis of particle eigenstates.
Thus, for an operator,
${\cal O}(x, \tau)$, we write the spectral decomposition schematically
($\tau$ denotes imaginary time, and $T$ time ordering)
\footnote{The sum over $s_n$ is meant to include integrals
over the momenta of all particles.}
\vbox{
\begin{eqnarray}\label{eIiii}
G^{\cal O}_T(x,\tau) &=& \langle 0|T\bigl (
{\cal O}(x,\tau){\cal O}^{\dagger}(0,0)
\bigr)|0\rangle \cr
&=& \hskip -.1in \sum_{n=0}^{\infty}
\sum_{s_n}
\langle 0|{\cal O}(x,0)|n;s_n\rangle e^{- \tau E_{s_n}}\times\cr
&& \hskip .35in \langle n;s_n|{\cal O}^{\dagger}(0,0)|0\rangle,
\qquad (\tau>0), 
\end{eqnarray}}
where $E_{s_n}$ is the energy
of the eigenstate, $|n;s_{n}\rangle$.
In an integrable model the matrix elements of the physical operator between
the vacuum and the exact eigenstates can in principle be computed
exactly from the 2-body S-matrix.
However the calculation of these matrix elements,
as well as the evaluation
of the sums/integrals
$\sum_{s_n}$, becomes increasingly cumbersome as the particle
number $n$ becomes large,
so that the full expression for the correlation function
cannot be evaluated in closed form.
Often, however, a truncation of the sum at the level of
two or three particle states already provides a good approximation
to the full correlation
function \cite{mus,les,lec,delone,deltwo,card}.
On the other hand, this truncation is no longer necessary
in a massive theory, 
if one considers the corresponding spectral function.
Only eigenstates with a fixed
energy, $\omega$, contribute to the spectral function:

\widetext
\top{-2.8cm}
\begin{eqnarray}\label{eIiv}
&&-{1 \over \pi} {\rm Im} G^{\cal O}_{T}(x, -i\omega + \delta ) =
\sum_{n=0}^{\infty}\sum_{s_n} \bigg\{ 
\langle 0|{\cal O}(x,0)|n;s_n\rangle
\langle n;s_n|{\cal O}^{\dagger}(0,0)|0\rangle  \delta(\omega- E_{s_n})\cr
\cr
&& \hskip 2.1in - \epsilon \langle 0|{\cal O}^\dagger(0,0)|n;s_n\rangle
\langle n;s_n|{\cal O}(x,0)|0\rangle  \delta(\omega + E_{s_n})\bigg\},
\end{eqnarray}
\bottom{-2.7cm}
\narrowtext
\noindent where $\epsilon = \pm$ 
for fields, $\cal{O}$, that are bosonic/fermionic.

Since in a massive theory the creation of an extra particle
in the intermediate exact eigenstate costs a finite
amount of energy, the sum in \ref{eIiv} is finite.
For example, when $\omega$ is smaller than the energy
of all three-particle states (i.e. when $\omega$ is below
the three-particle threshold), then only the form
factors with one and two particles $(n=1,2)$ have to be determined.
This is what we have done in this
paper for all the spectral functions we computed.
Thus, the results obtained from the form factor method
for spectral functions are exact in massive theories.

The results for the physical quantities we compute, namely the
optical conductivity, the single particle spectral function, 
and the tunneling
$I-V$ curve, are summarized in Section 3.  The basic
features of these results are as follows.
The optical conductivity, ${\rm Re}[\sigma (\omega)]$
(so called because it would be
measured in a reflectivity experiment) has two notable
features: a delta function peak at $\omega = \sqrt{3}m$ corresponding
to an excitation of one of the 28 rank two tensorial bosonic particles,
and a continuum of two particle states beyond $\omega = 2m$
(see Figures 1 and 2).  In a free theory, the
van-Hove square-root singularity in the density of states at the 
two particle threshold would lead ${\rm Re}[\sigma (\omega)]$ to behave as
$1/\sqrt{\omega - 2m}$ 
\footnote{This is the result found in \cite{lin}.  In this work
the optical conductivity is computed in the large $N$ limit of $SO(2N)$,
in effect an RPA approximation.  In this limit the theory consists of
four massive but non-interacting Dirac fermions.}.
However the current
matrix elements vanish at threshold changing the behaviour
to $\sqrt{\omega -2m}$.  This vanishing is ultimately the result
of generic interactions becoming strongly
renormalized at low 
energies \cite{leon}.
However its exact behaviour as one moves away from threshold 
can only be extracted through integrability.

As the $SO(8)$ Gross-Neveu model characterizes the low energy 
behaviour of the
Hubbard ladders/carbon nanotubes, 
we are able to describe the single particle 
spectral function only near the Fermi points.  Here we find
two principal features.  There is a sharp peak
(infinitely so at $T=0$) describing
the single electron contribution to the spectral function.  As
the system is interacting we expect to find a continuum of
higher particle contributions
to the spectral function.  In physical terms these higher particle
contributions would take the form of neutral particle-hole excitations
together with an electron.  In the Gross-Neveu language said 
excitations take
the form of a kink (akin to the electron) and a fundamental
Gross-Neveu fermion (akin to the particle hole excitation).
The threshold of this
two-particle contribution occurs at an energy, $\omega=2m$.

One of the quantities most easily measured in experiments
on nanotubes is the tunneling density of states. This
quantity can be measured by tunneling from a metallic
lead, such as an STM tip, into the nanotube.  When applying
a voltage bias, a current will flow. This current is
a non-linear function of the applied bias.
To lowest non-vanishing order 
in the tunneling matrix element, this current
can be
related to the single-particle spectral function
at energy $\omega=V$ (using standard reasoning). 
This relationship is non-perturbative
in the voltage $V$.
Differentiating the so-obtained non-linear current-voltage
characteristic, the differential conductance for tunneling
into the nanotube is easily
seen to equal the spectral function
of the single-particle Greens functions. Physically,
it is a `spectroscopic probe', or `measure'
of the density of states in the fully  interacting nanotube 
at energy $V$, even though it is not equal to it
as the spectral function is equal to the
density of states multiplied by the form-factor matrix elements.
A look at Figure 5 reveals an
interesting novel feature, entirely due to the
interactions, and not visible in a band-semiconductor:
the differential  tunneling conductance develops
a jump of a finite magnitude at the
two-particle threshold. This might have been expected on physical
grounds, since at this energy extra  states
suddenly become
available for current transport.  Nevertheless, is it
by no means obvious that this would correspond to a
discontinuity in the tunneling density of states;
the two-particle channel could also open up gradually.
This depends on the detailed behaviour of the matrix elements
near the threshold, which we have computed exactly from integrability. 
Our analysis shows 
that these matrix elements behave in such a way that
there is a finite jump.

The one and two particle form factors
have not in general been computed before
for the $SO(8)$ Gross-Neveu model.  The sole exception is
the matrix elements of the current operators with the fundamental
fermions \cite{weisz}.
The calculation of the remaining form factors,
and in particular those involving the kinks, makes up the technical
part of this paper.  To compute the form factors, we employ a series
of algebraic constraints that arise from consistency
with known two body scattering matrices, unitarity, Lorentz invariance, and
braiding relations.  The first three are common in form factor
calculations (see for example \cite{delone,deltwo,card,mus,les,lec,weisz}).
The latter is perhaps more unusual \cite{deltwo,smirbook,zam}.  
Braiding relations arise as generalized commutators
of fields, $\psi^a$:
\begin{equation}\label{Iv}
\psi^a (x,t)\psi^b (y,t) = R^{ab}_{cd}\psi^d (y,t) \psi^c (x,t), ~~~ x>y,
\end{equation}
where $R^{ab}_{cd}$ is termed the braiding matrix.
They are both indicative of the non-locality
of the fields for which we compute form factors and of exotic
symmetries.  Non-trivial braiding relations are tied to
the existence of quantum group symmetries \cite{smir}.
In this case, the relevant quantum group symmetry
is an $SO(8)$ Yangian.  This is similar to the situation
found in the well-studied sine-Gordon model at its SU(2) point
\cite{double},
where an SL(2) Yangian symmetry is present.

The $SO(8)$ Yangian is the operative symmetry of the model.
The particles in the theory are thus organized in terms of
its finite dimensional representations and not $SO(8)$'s.  There is 
however a large degree of correspondence between the two sets of 
representations.  
The three fundamental
eight dimensional representations of $SO(8)$ are also irreducible 
representations
of the Yangian.  However the 28 dimensional second rank 
anti-symmetric tensor 
representation does not 
appear as a representation of the Yangian.  Rather under the Yangian
it is combined with the one-dimensional scalar representation
into 
a single 29 dimensional representation\footnote{We would
like to thank N. MacKay for stressing this fact to us.}.  
Thus, unsurprisingly the scalar particle together with the 28 
particles of the
anti-symmetric tensor share the same mass, $\sqrt{3} m$.

The paper is organized as follows.  In Section 2, we review the arguments
given by Lin, Balents and Fisher \cite{lin} showing
how the
various massive phases of the Hubbard
ladders/nanotubes 
are related to an $SO(8)$ Gross-Neveu model.  In doing so
we establish notation.  In Section 3, as stated previously, we 
summarize our results for
the optical conductivity, the single-particle 
spectral function,
and the tunneling I-V curve.  Sections 4 and 5 are devoted to
computing the form factors.  For readers uninterested in the details
of this analysis, the last part
of Section 5 provides a summary of results.  Specifically, Section
4 reviews the S-matrices and attendant group theory for the
$SO(8)$ Gross Neveu model,
while Section 5 presents the actual form factor computations.

\section{Hubbard Ladders to $SO(8)$ Gross-Neveu}

Here we will briefly review the connection between Hubbard ladders
(and related armchair carbon nanotubes)
and the $SO(8)$ Gross-Neveu model developed in \cite{lin}.
Specifically, we summarize the map between the two
models, and interpret the excitation spectrum and fields of
the 
$SO(8)$ Gross-Neveu model
in terms of the original ladder model.

\subsection{Map: D-Mott Hubbard Ladders to $SO(8)$ Gross-Neveu}

\newcommand{\ccr}{c^\dagger_{Rj\alpha}}
\newcommand{\car}{c_{Rj\alpha}}
\newcommand{\ccl}{c^\dagger_{Lj\alpha}}
\newcommand{\calf}{c_{Lj\alpha}}

The weakly interacting Hubbard ladder has five different phases, 
one massless
and four massive \cite{lin}.  The four massive phases, D-Mott (a spin liquid
with approximate short-range $d$-wave pairing symmetry), S-Mott
(a spin liquid with approximate short-range $s$-wave pairing  symmetry),
spin-Peierls (SP) (electrons are dimerized along the legs of the ladder),
and a phase with charge density wave order (CDW), correspond to various
combinations of attractive and repulsive interactions.  We first focus
upon the D-Mott phase which is characterized by generically repulsive
interactions.  Once we discuss the D-Mott phase in detail, we will
return to the other three massive phases and demonstrate that each has a
distinct $SO(8)$ symmetry \cite{lin}.  In the massless phase, 
termed C2S2 (i.e. two
charge bosons and two spin bosons),
interactions are irrelevant, and so the phase has
two trivial independent $SO(8)$ chiral symmetries.
For this reason we will not consider it here.

We follow \cite{lin} in relating the D-Mott Hubbard ladder to an 
$SO(8)$ Gross-Neveu model.  We begin
with non-interacting electrons hopping on a ladder:
\begin{eqnarray}\label{eIIi}
H_0 = -\sum_{x,\alpha }
\bigl(
&& ta^\dagger_{1\alpha}(x+1)a_{1\alpha}(x)
+ ta^\dagger_{2\alpha}(x+1)a_{2\alpha}(x) \cr
&& ~~~~~ + t_{\perp}a^\dagger_{1\alpha}(x)a_{2\alpha}(x) + {\rm h.c.}
\bigr).
\end{eqnarray}
Here the $a_l$/$a_l^\dagger$ are the electron annihilation/creation
operators for the electrons on rung $l$ of the ladder, $x$ is a discrete
coordinate along the ladder, and $\alpha = \uparrow ,\downarrow$
describes electron spin.  $t$ and $t_\perp$ describe respectively
hopping between and
along the ladder's rung.

The first step in the map is to reexpress the $a$'s of $H_0$ in terms
of bonding/anti-bonding pairs:
\begin{equation}\label{eIIii}
c_{j\alpha} = {1\over \sqrt{2}}(a_{1\alpha} + (-1)^j a_{2\alpha}).
\end{equation}
With this transformation, the Hamiltonian can be diagonalized in
momentum space in terms of two bands.  Working at half filling, 
particle-hole
symmetry dictates that the Fermi velocities, $v_{Fj}$, of the two bands,
$j=1,2$, are equal.  As we are interested in the low energy behaviour of
the theory, the $c_{j\alpha}$'s are
linearized about the Fermi surface, $k_{Fj}$:
\begin{equation}\label{eIIiii}
c_{j\alpha} \sim \car e^{ik_{Fj}x} + \calf e^{-ik_{Fj}x},
\end{equation}
where $L,R$ corresponding to the right and left moving modes about
the Fermi surface.
With this $H_0$ becomes,
\begin{equation}\label{eIIiv}
H_0 = v_F \int dx \sum_{j\alpha}
\bigl [
 \ccr i\del_x \car - \ccl i\del_x \calf 
\bigr ].
\end{equation}
The next step is to bosonize the $c's$:
\begin{equation}\label{eIIv}
c_{Pj\alpha} = \kappa_{j\alpha} e^{i\phi_{Pj\alpha}},
~~~~ {\rm P = +,- = R,L}.
\end{equation}
Here $\kappa_{j\alpha}$ are Klein factors satisfying
\begin{equation}\label{eIIvi}
\{\kappa_{j\alpha},\kappa_{i\beta}\} = 2 \delta_{ij}\delta_{\alpha\beta}.
\end{equation}
In terms of these four Bose fields, four new Bose fields are defined
(effectively separating charge and spin):
\begin{eqnarray}\label{eIIvii}\nonumber\cr
\phi_{P1} &=& {1\over 2}(\phi_{P1\uparrow} + \phi_{P1\downarrow} +
\phi_{P2\uparrow} + \phi_{P2\downarrow}) ;\cr\cr
\phi_{P2} &=& {1\over 2}(\phi_{P1\uparrow} - \phi_{P1\downarrow} +
\phi_{P2\uparrow} - \phi_{P2\downarrow}) ;\cr\cr
\phi_{P3} &=& {1\over 2}(\phi_{P1\uparrow} - \phi_{P1\downarrow} -
\phi_{P2\uparrow} + \phi_{P2\downarrow}) ;\cr\cr
\phi_{P4} &=& {P\over 2}(\phi_{P1\uparrow} + \phi_{P1\downarrow} -
\phi_{P2\uparrow} - \phi_{P2\downarrow}) .\cr
\end{eqnarray}\begin{equation}\end{equation}
Note that $\phi_{P4}$ has a relative sign between the right and left movers.
This sign effectively masks the $SO(8)$ symmetry of the original 
Hamiltonian.
Without the sign change, \ref{eIIvii} is no more than a triality transformation
(see Section 2.4).
If we then refermionize with these new bosons, i.e.,
\begin{eqnarray}\label{eIIviii}
\Psi_{Pa} &=& \kappa_a e^{i\phi_a}, ~~~~ {\rm a = 1,...3}\cr\cr
\Psi_{P4} &=& P\kappa_4 e^{i\phi_4},\end{eqnarray}
where the Klein factors are given by
\begin{equation}\label{eIIix}
\kappa_1 = \kappa_{2\uparrow}, ~~~ \kappa_2 = \kappa_{1\uparrow}, ~~~
\kappa_3 = \kappa_{1\downarrow}, ~~~ \kappa_4 = \kappa_{2\downarrow},
\end{equation}
we find for the free Hamiltonian,
\begin{equation}\label{eIIx}
H = \int dx \sum_a 
\bigl [
\Psi^\dagger_{aL}i\del_x \Psi_{aL} -
\Psi^\dagger_{aR} i\del_x \Psi_{aR} 
\bigr ],
\end{equation}
where the Fermi velocity, $v_F$, has been set to 1.

The point to the bosonization, change of basis, and refermionization only
becomes apparent when one considers interactions.  The discovery
in \cite{lin} was that if one writes down a
generic set of weak, left-right, repulsive 
interactions between electrons on the rungs,
expresses these interactions in terms of the refermionized fermions, and
then allows the couplings to flow under an RG, one finds that the
interacting
Hamiltonian is attracted to,
\begin{equation}\label{eIIxi}
H_{int} =
g \big[\sum^4_{a=1} 
(i\Psi^\dagger_{aL}\Psi_{aR} - i\Psi^\dagger_{aR}\Psi_{aL})\big]^2 .
\end{equation}
This is, of course, $H_{int}$ for the $SO(8)$ Gross-Neveu model.

It will sometimes prove convenient to recast the theory in terms of 
Majorana fermions, $\psi_{aP}$.  In terms of the Dirac fermions, 
$\Psi_{aP}$,
they are given
by
\begin{equation}\label{eIIxii}
\Psi_{aP} = {1\over \sqrt{2}}(\psi_{2a,P} + i\psi_{2a-1,P}),
\qquad (a=1,...,4).
\end{equation}
In this basis, $H_{int}$ can be recast as
\begin{equation}\label{eIIxiii}
H_{int} = g G^{ab}_R G^{ab}_L, \qquad (a>b = 1,...,8),
\end{equation}
where $G^{ab}_P = i \psi_{aP}\psi_{bP}$ is one of the 28
$SO(8)$ Gross-Neveu
currents.

\subsection{D-Mott Excitations in Gross-Neveu}

The Gross-Neveu $SO(8)$ model has an exceedingly rich spectrum.  There are
24 fermionic particles of mass $m$ organized into one eight dimensional
vector representation and two eight dimensional spinor representations.
We denote the particles of the vector representation by $A_a$,
$a = 1,\ldots ,8$.  The $A_a$'s are the Majorana fermions of \ref{eIIxii}.
The kink particles, in turn, will be denoted by $A_\alpha$.  Here $\alpha$
is of the form $\alpha = (\pm 1/2, \pm 1/2,\pm 1/2,\pm 1/2)$ and so takes
on 16 values.  These 16 particles decompose into the two eight-dimensional
spinor representations.  This is discussed in more detail in Section 4.

Beyond the eight dimensional representations, there are
$29$ bosonic particle states of mass $\sqrt{3} m$,
transforming as a  rank-two tensor of dimension $28$
and a singlet.  Together they form a representation of the $SO(8)$ Yangian
symmetry.  
These particles can be thought of as
bound states of either two kinks or two fundamental fermions.

As $SO(8)$ is a rank 4 algebra, the $SO(8)$ Gross-Neveu 
model has four Cartan
bosons (i.e. the $\phi_{Pa}$ , $a = 1, \ldots 4$) and so its excitations
are characterized by four
quantum numbers, $N_i$, $i = 1,\ldots ,4$.
With the Majorana fermions, the combination
\begin{equation}\label{eIIxiv}
A_{2a} \pm iA_{2a-1},
\end{equation}
carries quantum number $N_a=\pm  1$, $N_b=0, b \not =  a$. 
The quantum numbers carried by the
kinks $A_\alpha$ are directly encoded in $\alpha$.
If $\alpha = (a_1,a_2,a_3,a_4)$, $a_i=\pm 1/2$, the $A_\alpha$ carries
the quantum numbers, $N_i = a_i$.  The quantum numbers carried by the rank
two tensor states can be directly deduced from the particles forming the
bound state.  As we will always think of the bound states in this way,
we will not list their quantum numbers directly.

The last thing needed in the section is to identify the relationship 
between the quantum numbers, $N_i$, and the physical quantum numbers of the
system, the
z-component of spin, $S_z$, the charge, $Q$, the difference in
z-component of spin
between the two bands, $S_{12}$, and the ``relative
band chirality'', $P_{12}$, defined as
$P_{12} = N_{R1} - N_{L1} - N_{R2} + N_{L2}$, where $N_{Pj}$ is the
number electrons in band $j$ with chirality $P$.
In \cite{lin} it was found:
\begin{eqnarray}\label{eIIxv}\nonumber\cr
(N_1=1,0,0,0) &\leftrightarrow& (Q=2, S_z = 0, S_{12} = 0, P_{12} = 0);\cr
\cr
(0,N_2=1,0,0) &\leftrightarrow& (Q=0, S_z = 1, S_{12} = 0, P_{12} = 0);\cr
\cr
(0,0,N_3=1,0) &\leftrightarrow& (Q=0, S_z = 0, S_{12} = 1, P_{12} = 0);\cr
\cr
(0,0,0,N_4=1) &\leftrightarrow& (Q=0, S_z = 0, S_{12} = 0, P_{12} = 2).
\end{eqnarray}
\begin{equation}\end{equation}
With this assignment, we can see that the vector representation of 
fundamental
fermions corresponds to states of two electrons in the original 
formulation.
For example, the fermion $A_2 \pm iA_1$ carries charge $\pm 2$
and no spin (the cooperons),
and the fermion $A_4 \pm iA_3$ carries spin, $S_z = 1$,
and no charge (the magnons). 
This makes concrete the earlier comment that the change of basis of bosons
in \ref{eIIvii} affects a charge-spin separation.
The spinor representations, the kinks, in turn correspond to single 
particle
excitations as their quantum numbers are combinations of $N_i/2$.

\subsection{D-Mott Gross-Neveu Fields}

In this section we make contact between the fields of the 
$SO(8)$ Gross-Neveu model
and the original fields of the Hubbard ladders.  As we have 
already discussed,
the fundamental (Dirac)
fermions of the vector representation are given by
\begin{eqnarray}\label{eIIxvi}
\Psi_{aP} &=& \kappa_a e^{i\phi_{aP}}, \cr
\Psi_{aP} &=& P\kappa_a e^{i\phi_{aP}} ,\end{eqnarray}
and carry quantum numbers corresponding to two electronic excitations.
However the $\Psi_{aP}$ are fermionic, whereas such excitations are
bosonic.  As such, $\Psi_{aP}$ are not simply related to a fermionic
bilinear of the original electrons but must be a fermion bilinear
multiplying some non-local field (a Jordan-Wigner string).  As we
will not compute correlators involving such fields in this paper, we
will not elaborate upon this (see \cite{lin}).

As discussed previously, the kinks correspond to single particle 
excitations.
Thus we expect to find that the kink fields are related to the original
electron operators.  This is true in part.  There are 32 kinks in total
(counting both left and right movers), but only sixteen electron
operators, the $c$'s and $c^{\dagger}$'s (four for each of the 
four Fermi points).  So
we expect only $1/2$ of the kinks to correspond to actual electron 
operators.

We represented the fundamental fermions in terms of the four Cartan bosons.
There is a corresponding representation for the kink fields:
\begin{equation}\label{eIIxvii}
\psi_{\alpha P} \sim e^{i\alpha \cdot \bar{\phi}_P} ,
\end{equation}
where $\bar{\phi} = (\phi_1 , \phi_2 , \phi_3 ,\phi_4 )$.
The kink fields that then correspond to the electron operators
$c$'s are as follows:
\begin{eqnarray}\label{eIIxviii}\cr
c_{R1\uparrow} &\sim& e^{i(\phi_{1R} + \phi_{2R} + \phi_{3R} + \phi_{4R})/2};
\cr
c_{R2\uparrow} &\sim& e^{i(\phi_{1R} + \phi_{2R} - \phi_{3R} - 
\phi_{4R})/2} ;\cr
c_{R2\downarrow} &\sim& e^{i(\phi_{1R} - \phi_{2R} + \phi_{3R} - 
\phi_{4R})/2} ;\cr
c_{R1\downarrow} &\sim& e^{i(\phi_{1R} - \phi_{2R} - \phi_{3R} + 
\phi_{4R})/2} ;\cr\cr
&& {\rm (even~chirality)}\cr
\cr\cr
c_{L1\uparrow} &\sim& e^{i(\phi_{1L} + \phi_{2L} + \phi_{3L} - 
\phi_{4L})/2} ;\cr
c_{L2\uparrow} &\sim& e^{i(\phi_{1L} + \phi_{2L} - \phi_{3L} + 
\phi_{4L})/2} ;\cr
c_{L2\downarrow}
&\sim& e^{i(\phi_{1L} - \phi_{2L} + \phi_{3L} + \phi_{4L})/2} ;\cr
c_{L1\downarrow}
&\sim& e^{i(\phi_{1L} - \phi_{2L} - \phi_{3L} - \phi_{4L})/2} ; \cr\cr
&& {\rm (odd~chirality)} .
\end{eqnarray}
With hermitian conjugates, this totals to sixteen fields.
The $\sim$ sign is meant to indicate that these equivalences hold up to
Klein factors.  The $c_{Pj\alpha}$'s, of course, are fermionic.  However
the kink fields as defined are not.

The last set of fields that are of concern to us are the currents.  The
electric current of the ladder has the lattice representation
\begin{equation}\label{eIIxix}
J \sim -i \sum_{l\alpha}
\biggl [
 a^\dagger_{l\alpha}(x)a_{l\alpha}(x+1)
- a^\dagger_{l\alpha}(x+1)a_{l\alpha}(x) 
\biggr ],
\end{equation}
where we have summed over the contribution coming from each spin ($\alpha$)
and each leg ($l$) of the ladder. 
Taking the continuum limit, $J$ equals, in Gross-Neveu language, 
\begin{equation}\label{eIIxx}
J \sim i \sin k_{F1} ~\del_t \phi_1 \sim G_{12} ,
\end{equation}
where $G_{12}$ is one of the $SO(8)$ currents discussed in \ref{eIIxiii}.

\subsection{Other Massive Phases}

By adjusting the signs of interaction couplings,  three other massive
phases, S-Mott, SP, and CDW, can be obtained \cite{lin}. 
 Each of these phases
is characterized by a distinct $SO(8)$ symmetry.  However they share
a common $SO(5)$ subgroup.  The $SO(5)$ symmetry is the one-dimensional
analog of the $SO(5)$ symmetry recently proposed by S.C. Zhang as
a means to unify antiferromagnetism and superconductivity.
Each of these
new $SO(8)$ symmetries is readily expressible in terms of the D-Mott
$SO(8)$ symmetry through considering the four 
Cartan bosons $\Phi_{Pa}$ \cite{lin}.
As the first two Cartan generators
belong to the $SO(5)$ subgroup they, however, remain
unchanged.  It is the latter two bosons that are affected.
For completeness we review  each of new phases in turn.

\vskip .1in
\noindent {\it S-Mott:}
\vskip .1in

The defining four Cartan bosons of the S-Mott phase, $\phi^S_{Pa}$, 
are related
to those of the D-Mott phase via
\begin{eqnarray}\label{eIIxxi}
\phi^S_{Pa} &=& \phi_{Pa} , ~~~ a = 1,2,3 \cr\cr
\phi^S_{P4} &=& \phi_{P4} + P\pi/2 .\end{eqnarray}
The sole difference between the $SO(8)$ algebra of the S-Mott phase
and that of the D-Mott phase is a shift of $\phi_{Pa}$ by $P\pi/2$.
This shift forces a sign change in the pair field correlator, changing
the symmetry from $d$-wave to $s$-wave.  The shift, 
however, does not change
the excitation spectrum or its attendant assignment of quantum numbers,
nor does it change field assignments beyond a phase.

\vskip .1in
\noindent {\it Spin-Peierls (SP):}
\vskip .1in

The defining four Cartan bosons of the SP phase, $\phi^{SP}_{Pa}$, are
related to those of the D-Mott phase via
\begin{eqnarray}\label{eIIxxii}
\phi^{SP}_{Pa} &=& \phi_{Pa} , ~~~ a = 1,2,4 \cr\cr
\phi^{SP}_{P3} &=& P\phi_{P3}  ;\end{eqnarray}
that is, the sole change is to flip the sign on $\phi_{L3}$.  The effect
of this sign change is two-fold.  The quantum number associated with
the third Cartan boson is now defined to be
\begin{eqnarray}\label{eIIxxiii}
N_3 = S_{12} &=& {1\over 2}(N_{R\uparrow 1} - N_{R\downarrow 1} -
N_{R\uparrow 2} + N_{R\downarrow 2}) -\cr\cr
&& {1\over 2}(N_{L\uparrow 1} - N_{L\downarrow 1} -
N_{L\uparrow 2} + N_{L\downarrow 2}),
\end{eqnarray}
where $N_{Pj\sigma}$ is the number of electron of chirality $P$ in band
$j$ with spin $\sigma$, and so
$N_3$ is the relative right-moving spin between the two bands minus
the relative left-moving spin between the two bands.
Moreover
the left moving electrons, $c_L$, are now to be identified with kinks
of even chirality as opposed to
the odd chirality kinks for the D-Mott phase (see \ref{eIIxviii}).

\vskip .1in
\noindent {\it Charge Density Wave (CDW):}
\vskip .1in

The defining four Cartan bosons of the CDW phase, $\phi^{CDW}_{Pa}$,
are given by
\begin{eqnarray}\label{eIIxxiv}
\phi^{CDW}_{Pa} &=& \phi_{Pa} , ~~~ a = 1,2 \cr\cr
\phi^{CDW}_{P3} &=& P\phi_{P3}  \cr\cr
\phi^{CDW}_{P4} &=& \phi_{P4}  + P\pi/2 .\end{eqnarray}
Because of the sign change of $\phi_{P3}$, we again, as with the SP
phase, have a redefinition
of $N_3$ and reassociation of the kinks with the physical electrons.
The shift of $\phi_{P4}$, as with the S-Mott phase, only leads to
phase multiplication of certain fields.

\subsection{Triality}

Of the $SO(2N)$ groups, 
$SO(8)$ possesses a special symmetry called triality that
rotates its three fundamental representations (the vector and the two
spinors) among one another.  It is unique to $SO(8)$ as only in this case
do the spinor and vector representations have the same dimension.
The symmetry is isomorphic to $Z_3$
(hence its name) and so has a generator, g, such that $g^3 = 1$.

To exhibit the action of $g$ upon the three representations, we consider
its behaviour on the four Cartan bosons, $\phi_{aP}$, introduced in
\ref{eIIvii}.  Under $g$ the $\phi_i$'s are linearly transformed as follows
\begin{eqnarray}\label{eIIxxv}
\phi_{1P} &&\rightarrow (\phi_{1P} + \phi_{2P} + \phi_{3P} + 
\phi_{4P})/2 ;\cr\cr
\phi_{2P} &&\rightarrow (\phi_{1P} + \phi_{2P} - \phi_{3P} - 
\phi_{4P})/2 ;\cr\cr
\phi_{3P} &&\rightarrow (\phi_{1P} - \phi_{2P} + \phi_{3P} - 
\phi_{4P})/2 ;\cr\cr
\phi_{4P} &&\rightarrow (\phi_{1P} - \phi_{2P} - \phi_{3P} + 
\phi_{4P})/2 .
\end{eqnarray}
Given that the four Cartan bosons, $\phi_{aP}$ are identified with the
four quantum numbers, $N_i$, $g$ acts to redefine them
correspondingly.

As the fermions are represented by $e^{i\phi_{aP}}$ and the kinks by
$e^{i\alpha\cdot\bar{\phi}_P}$, we can see the triality transformation
acts to take the fermions to the kinks of even chirality (i.e. the number
of components of $\alpha$ that are negative is even), the even kinks to
the odd kinks, and the odd kinks to the fermions.

We will use triality to fix some of the properties of the form factors in
Section 5.  Via triality, we can relate form factors involving fermionic
and kink fields and excitations, thus constraining them.

\section{Presentation of Results}

In this section we present the computations using form factors of the
optical conductivity, $\sigma$, the single particle spectral function,
$A(k,w)$, and a tunneling $I-V$ curve.  These results are generic to
any of the four massive $SO(8)$ phases of the Hubbard ladders/carbon
nanotubes.  The
optical conductivity is the same in each of the phases as they
share a common $SO(5)$ subalgebra that contains the conserved 
electric charge.
The computation of the spectral function and tunneling $I-V$ curve
is done with respect to generic kinks.  There is no need to specify
which exact kink one is working with as one obtains
identical results independent of the kink type (predominantly because
of the action of the $SO(8)$ symmetry).  Thus the results are independent
of the identification of the kinks with the physical electronic
excitations one makes in each of the massive phases.

In the Sections 4 and 5 that follow,
we lay out the calculation of the form factors.  For those
uninterested in the details, the bulk of these sections
may be skipped with only the results at the end of Section 5 referenced.

\subsection{Optical Conductivity}

In this section we consider the response of the ladder system to an
electric field polarized along the legs.  Apart from the
treatment in \cite{lin}, this problem
has been examined previously,
both theoretically \cite{scalp} and experimentally
\cite{tak}.  However
these two latter papers did not consider undoped
ladders at zero temperature.

In linear response, the optical conductivity is given by
\begin{equation}\label{eIIIi}
{\rm Re} \big[\sigma (\om ,k)\big]
= {\rm Im} \big[ {\Delta (\om ,k) \over \om}\big],
\end{equation}
where $\Delta$ is the current-current correlator
\begin{equation}\label{eIIIii}
\Delta (\om ,k) = \int dx d\tau e^{i\om \tau}e^{i x k} 
\langle T(J(x,\tau)J(0,0))
\rangle |_{\om \rightarrow -i\om + \delta}.
\end{equation}
$J$ is given by \ref{eIIxx},
\begin{equation}\label{eIIIiii}
J \sim G^{12}_1.
\end{equation}
To compute the correlator, $\tcor$, we insert a resolution of the identity
between the two $J's$, turning the correlator into a form factor sum.
We then have

\widetext
\top{-2.8cm}
\begin{eqnarray}\label{eIIIiv}
\tcor &=& \sum^\infty_{n=0}\sum_{a_1,\cdots ,a_n} \int {d\th_1\over 2\pi}
\cdots {d\th_n \over 2\pi} 
\langle G^{12}_1 (0)|A^\dagger_{a_1}(\th_1) \cdots
A^\dagger_{a_n}(\th_n) \rangle \cr
&& \hskip -1in \times \langle A_{a_n}(\th_n) \cdots
A_{a_1}(\th_1)| G^{12}_1 (0)\rangle
\exp \big(-|\tau| \sum^n_{i=1} m_{a_i}\cosh (\th_i)
+ ix\sum^n_{i=1} m_{a_i}\sinh (\th_i)\big),
\end{eqnarray}
\bottom{-2.7cm}
\narrowtext
\noindent where 
the first sum $\sum_n$ runs over the number of particles in the form
factor expansion and the second sum $\sum_{a_i}$ runs over 
the different particle
types.  We have also extracted the spacetime dependence of each term.

To compute this sum in its entirety is generally an intractable problem.
The usual solution is to truncate the sum at some $n$.  Here we will content
ourselves with a truncation at the two particle level:
\widetext
\begin{eqnarray}\label{eIIIv}
\tcor &=& \int {d\th_1 \over 2\pi} \langle G^{12}_1(0) | A^\dagger_{12}
(\th_1)\rangle
\langle A_{12}(\th_1) | G^{12}_1(0)\rangle \cr
&& ~~~~~~~~~~~~ \times \exp\big(-|\tau|\sqrt{3}m\cosh (\th_1)\big)
+ ix\sqrt{3}m\sinh (\th_1)) \cr
&& \hskip -.75in + \int {d\th_1 \over 2\pi} {d\th_2 \over 2\pi}
\exp\big(-|\tau|m(\cosh(\th_1)+\cosh(\th_2))
+ ixm(\sinh(\th_1)+\sinh(\th_2))\big)
\cr
&&
\times \big( \sum_{ab} \langle G^{12}_1(0)|A^\dagger_a(\th_1)A^\dagger_b
(\th_2)\rangle
\langle A_b(\th_2)A_a(\th_1)|G^{12}_1(0)\rangle \cr
&& +
\sum_{\alpha\beta} \langle G^{12}_1(0)|A^\dagger_\alpha (\th_1)
A^\dagger_\beta (\th_2)\rangle
\langle A_\beta(\th_2)A_\alpha(\th_1)|G^{12}_1(0)\rangle \big) .
\end{eqnarray}
\bottom{-2.7cm}
\narrowtext
\noindent The first term gives 
the single particle contribution to the correlation
function.  The only particle that contributes here is $A_{12}$,
denoting one of the particles
belonging to the rank 2 tensor multiplet.  At the two particle level a
variety of contributions are non-zero.  The second term in \ref{eIIIv} gives
the contribution of two Majorana fermions while the third term gives
the contribution of kinks with the same chirality.

As indicated in the introduction, this truncation of the form factor
sum is better than it may at first seem.  Because the correlator 
is evaluated
at zero temperature in a massive system, the higher order terms make
contributions only at higher energies, $\om$.  That is, the massiveness
of the system leads to particle thresholds.  
The next contribution comes from a three particle combination of
even kink/fermion/odd kink that carries mass $3m$.  Thus for $\om < 3m$,
this term gives no contribution to ${\rm Re} [\sigma (\omega,k)]$, 
for arbitrary $k$.
Hence our result for ${\rm Re} [\sigma (\omega ,k)]$, is exact 
for $\omega <3m$. 

In the case when $\om$ does exceed $3m$, we expect the higher particle
form factors to make only a small contribution to $\sigma (\om )$.
In practice, form factor sums have been found to be strongly convergent
for operators in massive theories \cite{mus,delone,deltwo,card}.
To obtain a good approximation to correlators involving such fields,
only the first few terms need to be kept.
Even in massless theories where there are no explicit
thresholds, convergence is good provided the 
engineering dimension of the operator matches its anomalous
dimension \cite{les,lec}.

Using the results for the form factors of Section 5, we can put everything
together and write down an expression for ${\rm Re} [\sigma (\om )]$:
\widetext
\top{-2.8cm}
\begin{eqnarray}\label{eIIIvi}
{\rm Re} [\sigma (\om )] &=& \delta(w-\sqrt{3m^2+k^2})
~({A_G\over m})^2 {2 \over 9}\sqrt{\pi \over 3}
{\Gamma (1/6) \over \Gamma (2/3)}
\exp\big[-2\int^\infty_0 {dx \over x} {G_c(x) \over s(x)} s^2 (x/3)\big]\cr
&& + \theta (w^2 - k^2 - 4m^2) {12 m^2 A_G^2 \over (\om^2 - k^2 - 3m^2)^2}
{\om {\sqrt{\om^2-k^2-4m^2}}\over (\om^2 - k^2)^{3/2}}  \cr
&& ~~~~~~~~~~~~~~~~~~~~~~~~~~~\times
\exp \big[ \int^\infty_0 {dx \over x}{G_c(x) \over s(x)}
(1-c(x)\cos ({x\th_{12}\over\pi}))\big] ,\end{eqnarray}
\bottom{-2.7cm}
\narrowtext
\noindent where $s(x) = \sinh (x)$, $c(x) = \cosh (x)$, and
\begin{equation}\label{eIIIvii}
\th_{12} = \cosh^{-1} \big[{\om^2-k^2-2m^2 \over 2m^2}\big].
\end{equation}
As indicated in Section 5, $A_G$ is an arbitrary constant
normalizing all current form-factors while $G_c (x)$ can be found
in (5.68) .

In Figure 1 we plot the real part of the optical 
conductivity for wavevector $k=0$.
There is no Drude weight as we are at zero temperature and zero doping.
We see there is an exciton type peak at $\om = \sqrt{3}m$ corresponding
to the single particle form factor contribution. 
The first vertical
dashed line marks out the beginning of the two particle
form factor contribution to the conductivity.
The onset of the two particle
contribution behaves as $\sqrt{\omega -2m}$ and not 
as $1/\sqrt{\omega -2m}$ as would
be expected in a free theory due to the divergence in the
density of states, the van-Hove singularity,
that generically occurs in one dimensional systems.  This singularity
is removed by the corresponding current matrix element which behaves
as $(\om-2m)$ with $\om \rightarrow 2m^+$ as is expected generically
because the low energy behaviour becomes strongly renormalized in
the presence of even weak interactions \cite{leon}.

The optical conductivity was computed 
in \cite{lin} using the large $N$ limit
of $SO(2N)$, or in an alternate language, an RPA approximation.
In such an approximation, the model becomes equivalent to a theory
of four massive, non-interacting Dirac fermions.  Hence \cite{lin} finds
that the van-Hove singularity is present.

The second vertical dashed line in Figure 1
at $3m$ marks the point where the three particle form factors would begin
to make a contribution.  Up to this point, the result is exact.  
We note that the three particle contribution is strictly a consequence of
interactions.  In $SO(8)$ language, two kinks of opposite chirality
together with a fermion will couple to the current operator.
In a free theory these different particles 
would not all exist and there would be
no three particle contribution.

If we were to compute the three particle contribution, three possibilities
present themselves.  
The three particle
density of states approaches a constant as $\om \rightarrow 3m^+$.  
If the corresponding matrix element vanishes
as $\om \rightarrow 3m^+$, the contribution opens up gradually,
leaving $\sigma (\om )$ continuous at $\om = 3m$.
If the three
particle matrix element also approaches a constant value as
$\om \rightarrow 3m^+$, the conductivity
is marked by a jump at $\om = 3m$.  But if the matrix element 
diverges in this limit, we expect a corresponding
divergence in the conductivity at $\om = 3m$.

\vskip .25in
\hskip -.4in 
\centerline{\psfig{figure=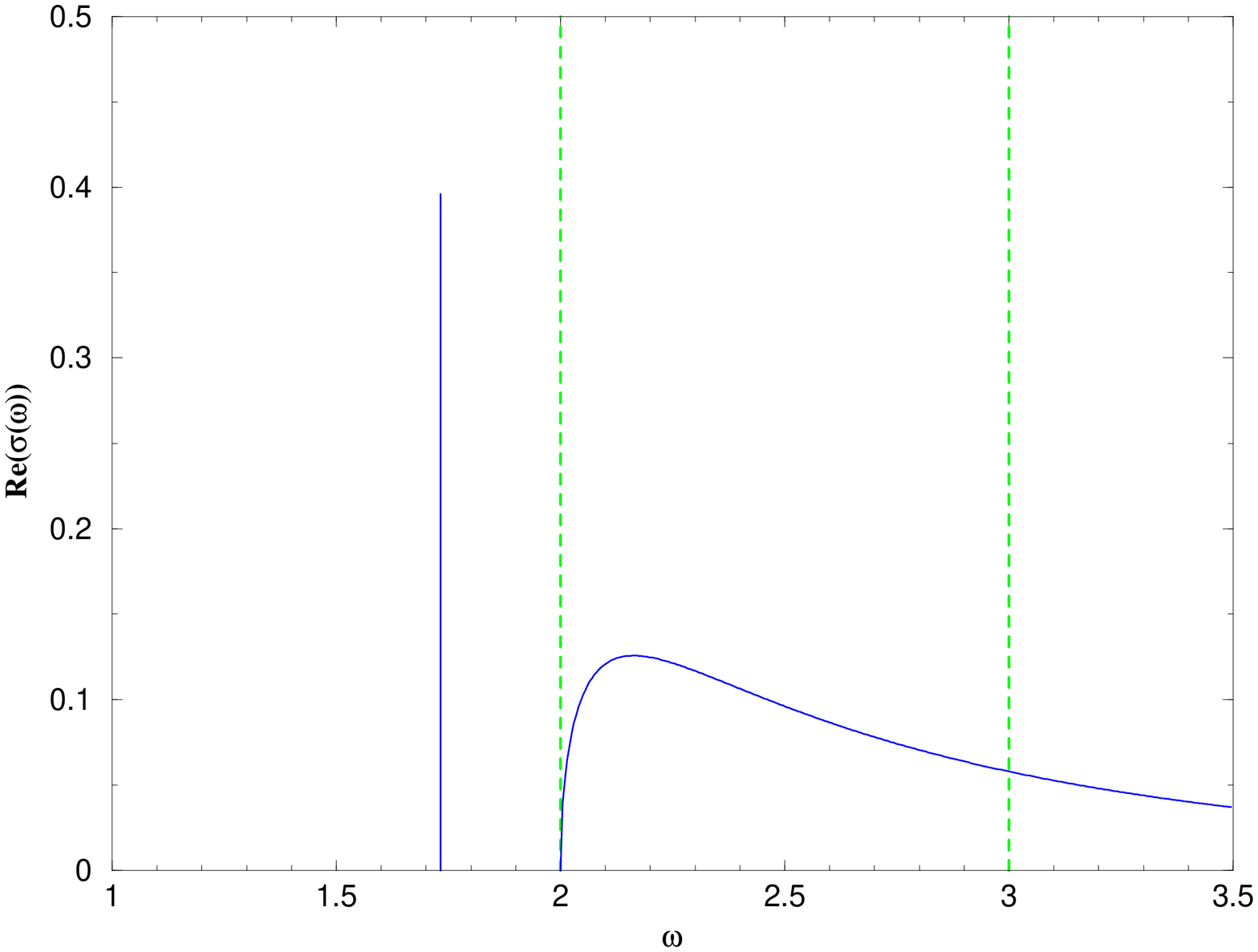,angle=-90,height=3.in,width=2.625in}}
\vskip .25in
\indent\vbox{\noindent\hsize 3in Figure 1: Plot of the optical 
conductivity at
wavevector $k=0$.}
\vskip .25in

In Figure 2 we plot the real  part of the optical 
conductivity for wavevector $k=\pi /3$.
As a result of the form of ${\rm Re}[(\sigma )]$  (see \ref{eIIIvi}), shifting
$k$ from $0$ to $\pi/3$ shifts the single particle form factor
contribution to $m\sqrt{\pi^2/9 + 3}$ and the onset of the
two particle contribution to $m\sqrt{\pi^2/9+4}$.  We see that
as $k$ grows, the contribution from the single particle moves
closer to the two particle threshold.

It can now be asked how perturbations to $SO(8)$ Gross-Neveu
will affect the computation
of the optical conductivity.  We consider this in the broadest terms by
focusing upon how the spectrum of $SO(8)$ Gross-Neveu is changed under a
perturbing term.  We 
do so through straightforward stationary perturbation theory, in
the same spirit that \cite{simo} treated the off-critical Ising model in
a magnetic field.  The most
general possible perturbation takes the form
\begin{equation}\label{eIIIviii}
H_{\rm pert} = \lambda G_{ab}G_{cd} ,
\end{equation}
where $G_{ab}$, $G_{cd}$ are $SO(8)$ currents (of unspecified chirality).
For such a perturbation it is necessary to consider degenerate perturbation
theory.  Thus in a given particle multiplet (for example, the fundamental
fermions in the vector representation), the perturbed energies arise
through diagonalizing the matrix
\begin{equation}\label{eIIIix}
M_{ij} = \langle A_i^\dagger (\theta ) H_{\rm pert} A_j (\theta )\rangle ,
\end{equation}
where here the index $i,j$ indicates the particles $A_i$,$A_j$ belong to
the multiplet of concern.
In the case that $G_{ab} = G_{cd}$, $M_{ij}$ is 
necessarily diagonal, i.e. nondegenerate perturbation theory is sufficient.
\vskip .25in
\hskip -.4in 
\centerline{\psfig{figure=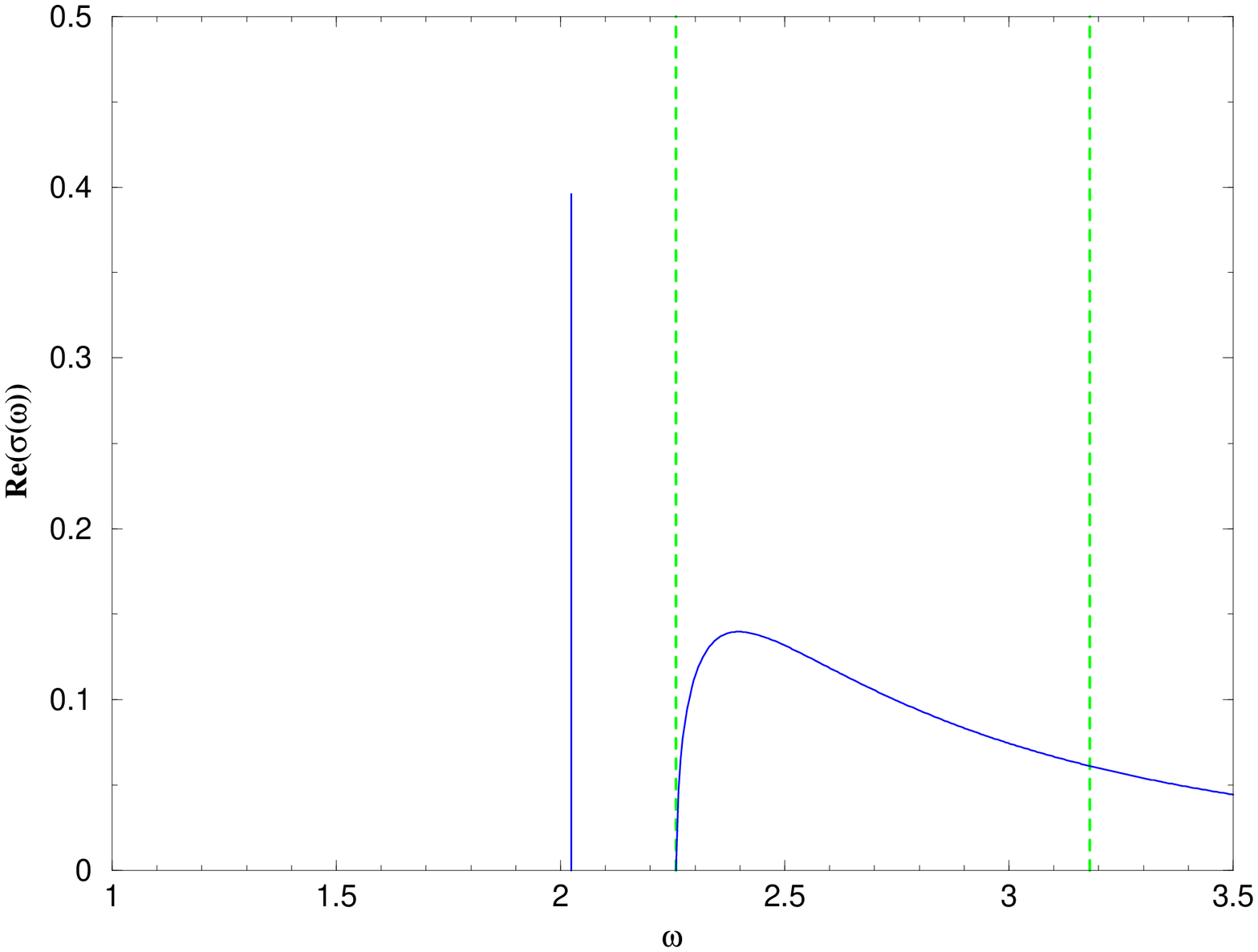,angle=-90,height=3in,width=2.625in}}
\vskip .25in
\indent\vbox{\noindent\hsize 3in Figure 2: Plot of the optical 
conductivity at
wavevector $k=\pi /3$.  This plot is exact up to $w/m = \sqrt{\pi^2/9 +9}$ 
at which point the three particle states begin to make a contribution.}
\vskip .25in

It is important to emphasize that this procedure can be handled in the
context of integrability.  The expression in \ref{eIIIix} is no more than
a form-factor which can readily be computed.  
Moreover as the theory is massive,
perturbation theory is well controlled.
We expect the unperturbed theory to describe all qualitative features
of the model while the perturbations to only introduce small quantitative
changes.

The consequences of a perturbation for the optical conductivity are
two-fold.  We expect the exciton peak (found, for 
example, in Figure 1 at $k=0$ and $\omega = \sqrt{3}$) 
to split.  In the unperturbed model
the peak results from a single rank two bosonic bound state coupling to the
current operator.  When the matrix
$M_{ab}$ above is diagonalized, this particular state should be 
mixed into many
others resulting in several states that couple to the current operator.
However we do not expect the functional forms of the exciton peaks 
to change: they should
remain delta functions.  They must do so provided the perturbation is
not so large as to push the exciton peak 
past the threshold of two particle
states where it then conceivably could decay.  As there is a gap between
the excitonic peak and the two particle threshold, this will not
happen for small perturbations.  Experimentally changes to the exciton
peak may
not be detectable.  Given that any experiment will be conducted at finite
temperature, the excitonic peak will be thermally broadened, perhaps
washing out any splitting of the original zero temperature peak.

We also expect the perturbation to affect the onset of
the two particle threshold, although in a less dramatic fashion.  
Like the unperturbed case, there will be several two particle contributions 
to the optical conductivity.  However unlike the unperturbed case,
the thresholds of the two particle contributions will not all occur
at $\omega = 2m$ but be distributed about this energy.  Thus the
two particle contribution is arrived at (approximately) by superimposing
several slightly shifted two-particle contributions similar to those 
found in Figures 1 and 2.  But given
the optical conductivity vanishes at threshold, the qualitative picture
remains effectively unchanged (i.e. the superimposed contributions will
appear nearly identical to the original picture).  
That the optical conductivity vanishes
at the two particle threshold is a result of the vanishing of the
relevant matrix element at threshold.  This should be robust under
perturbation as it is ultimately a consequence of the mere presence of
interactions and not some particular type of interactions.

\subsection{Single Particle Spectral Function}

In this section we
compute the  single-particle 
spectral function of the electrons of the  Hubbard ladder. 
To do so we first consider the correlator,
\begin{eqnarray}\label{eIIIx}
G (k_x,k_y,\tau) &=& 
\sum_{l=1,2} \int^\infty_{-\infty} dx ~e^{-ik_y l - ik_x x}\cr
&& \hskip .4in \times \lb T(a_{l\alpha}(x,\tau)a^\dagger_{l\alpha}(0,0))\rb .
\end{eqnarray}
Here $k_y$ takes on the values $0,\pi$.
We then define the particle/hole spectral functions, $A_{p/h}$, as follows:
\begin{eqnarray}\label{eIIIxi}
A_p({\bf k},\om ) + A_h(-{\bf k},-\om ) &=& \cr
&& \hskip -1in 
{\rm Im} \int^\infty_{-\infty} d\tau~ e^{-i\om \tau} G(k_x,k_y,\tau)
\big|_{\om\rightarrow -i\om + \delta} .
\end{eqnarray}
In keeping with \cite{lin}, we have not explicitly summed over spin, $\alpha$.

As described in Section 2,
electronic excitations around the Fermi point
correspond in the Gross-Neveu
language  to low energy excitations of kinks.  We thus expect to recast
the Greens function, $G$, above in terms of kink correlators.
This in fact can be done with the result,
\begin{eqnarray}\label{eIIIxii}
G (Pk_{Fi} + k,k_{yi},\tau) &=& \cr
&& \hskip -1in \int^\infty_{-\infty} dx e^{ikx}
\lb T(c_{Pi\alpha}(x,\tau)c^\dagger_{Pi\alpha}(0,0))\rb ,
\end{eqnarray}
where $i=1,2$ and $k_{yi} = (2-i)\pi$.  The $c$'s, the bonding-anti-bonding
electrons given in \ref{eIIii}, are in turn related to the various kinks
via \ref{eIIxviii}.
The Greens function on the r.h.s. of \ref{eIIIxii} is thus equal to 
\begin{equation}\label{eIIIxiii}
\lb T(c_{Pi\alpha}(x,\tau)c^\dagger_{Pi\alpha}(0,0))\rb = 
\langle T(\kappa_\alpha\psi_\pm^\alpha (x,\tau)
\kappa_{\bar{\alpha}}\psi_\pm^{\bar{\alpha}} (0,0)) \rangle.
\end{equation}
where $\alpha$ ($\bar{\alpha}$ being its charge conjugate) is 
the particular kink corresponding
to the Fermi point $(k_{Fi},k_{yi})$.
The $\kappa_\alpha$ are Klein factors included to ensure the $\psi^\alpha$
are anti-commuting.
Because of the $SO(8)$ symmetry together with its associated triality
symmetry, 
$\langle T(\kappa_\alpha\psi_\pm^\alpha (x,\tau)
\kappa_{\bar{\alpha}}\psi_\pm^{\bar{\alpha}} (0,0)) \rangle$ 
turns out to be independent of the type $\alpha$  of kink. 
It is only
sensitive to whether the kink field is right ($+$) or left ($-$) moving.

To compute this correlator, we again expand to the two lowest contributions:
\begin{eqnarray}\label{eIIIxiv}
\langle T(\kappa_\alpha\psi_\pm^\alpha (x,\tau > 0)
\kappa_{\bar{\alpha}}\psi_\pm^{\bar{\alpha}} (0,0)) \rangle
&=& \cr\cr
&& \hskip -1.5in \int^\infty_{-\infty} {d\th_1 \over 2\pi}
\langle \psi_+^\alpha(x,\tau) A^\dagger_{\bar{\alpha}}(\th_1) \rangle
\lb A_{\bar{\alpha}}(\th_1)\psi_+^{\bar{\alpha}}(0)\rb \cr\cr
&& \hskip -1.5in + \sum_{a\beta} \int^\infty_{-\infty} {d\th_1 \over 2\pi}
{d\th_2 \over 2\pi} 
\langle \psi_+^\alpha(x,\tau) A^\dagger_\beta (\th_2)
A^\dagger_a (\th_1) \rb \cr
&& \hskip -.45in\times 
\lb A_a(\th_1) A_\beta (\th_2) \psi_+^{\bar{\alpha}}(0)\rb .\end{eqnarray}
The first contribution, the one particle contribution, 
comes from the kink excitation,
$A_{\bar{\alpha}}$, destroyed
by the field, $\psi_\alpha$.  The second contribution, 
a two particle contribution,
arises from kinks, $A_\beta$, of opposite chirality to $A_\alpha$, and
Majorana fermions, $A_a$. 
(This reflects the group theoretical fact that
the tensor product of an $SO(8)$ spinor representation 
with an $SO(8)$ vector representation
gives the other $SO(8)$ spinor representation \cite{sla}.)  
The first contribution not included
is a bound state-kink pair.  
It begins to contribute at $\om = (1+\sqrt{3})m$.

From the form factor expressions from Section 5, we can then write
down the expression for the spectral functions, $A_{p/h}(\om ,k)$,
\widetext
\top{-2.8cm}
\begin{eqnarray}\label{eIIIxv}
A_p(\om ,Pk_{Fi}+k,k_{yi}) &=& A_h(\om ,-Pk_{Fi}+k,k_{yi}) 
= {\pi |c_P|^2 \over m} {\om + Pk \over \sqrt{k^2+m^2}}
\delta (\om - \sqrt{k^2+m^2}) \cr
&& + \theta (\om - \sqrt{k^2+4m^2})
{16 m^4 A_F^2 \over \om - P k }{1\over (\om^2-k^2-m^2)^2}
{1\over \sqrt{\om^2-k^2-4m^2}}\cr
&& ~~~~~~~~ \times\exp\big[\int^\infty_0{dx\over x}{G_f(x)\over s(x)}
(1-c(x)\cos ({x\th_{12}\over\pi}))\big],
\end{eqnarray} 
\bottom{-2.8cm}
\narrowtext
\noindent where $A_F$ 
is the (unspecified) normalization of the two particle
kink form-factor, $G_f(x)$ 
is given in (5.70), and $c_\pm$ is found in (5.73).
For $P=R=+$ (i.e. right-moving electrons/kinks), this function
is plotted in Figure 3.

\hskip -.5in\centerline{\psfig{figure=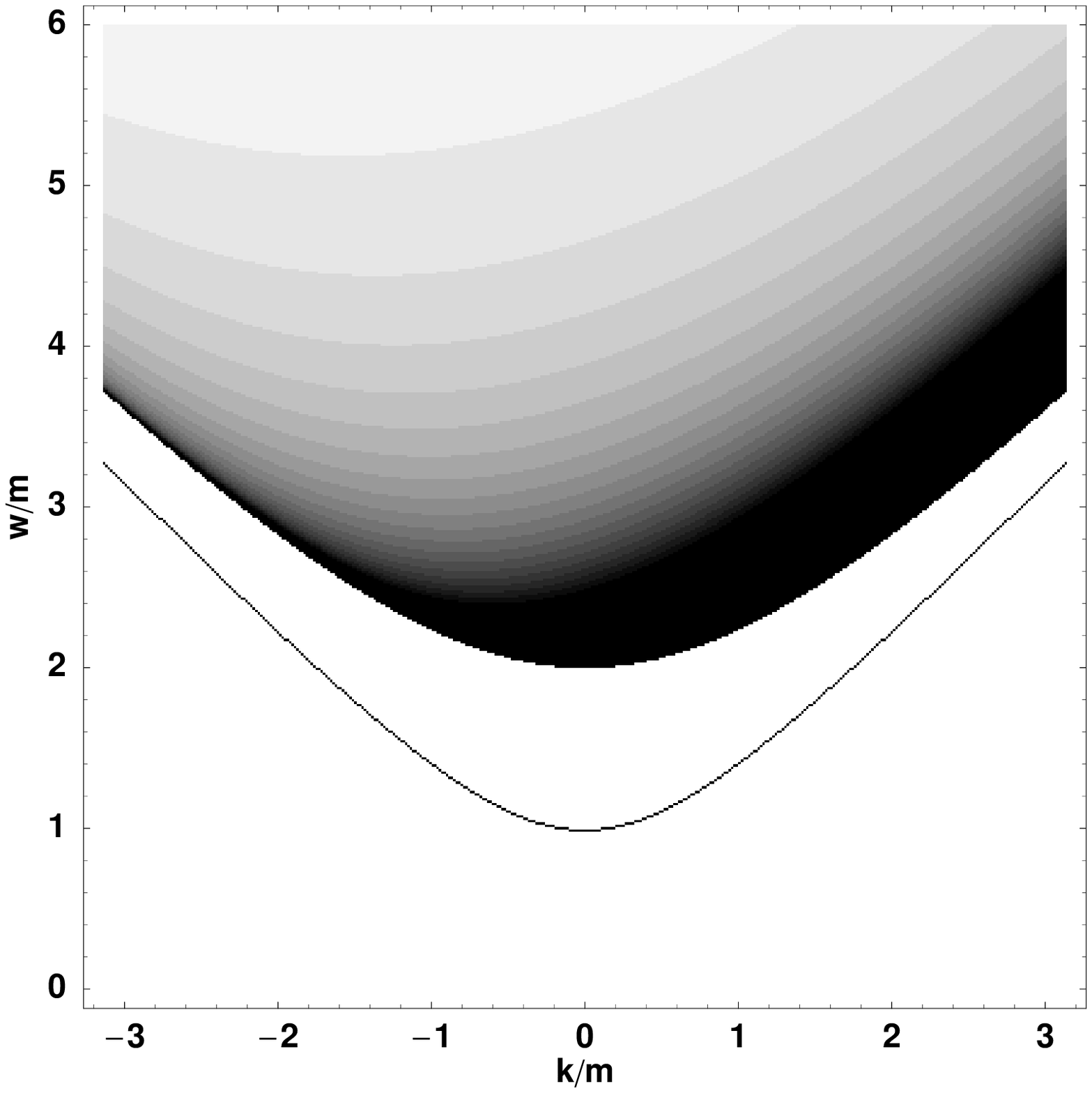,height=4in}}
\vskip -.75in
\indent\vbox{\noindent\hsize 3in Figure 3: Plot of the single particle 
spectral function for right moving kinks.  The more darkly shaded region
corresponds to greater spectral weight.}
\vskip .25in 

The parabolic line in Figure 3 arises from the single particle form
factor contribution, and represents the standard dispersion relation
of a particle of mass, $m$.  Above this curve comes the two particle
form factor contribution to the spectral function.  This contribution
is bounded by the curve, $\om = \sqrt{k^2 + 4m^2}$, and so the single
particle states do not cross into the two particle region.  As
can be seen from \ref{eIIIxv}, the two particle contribution opens up
at threshold with a square-root singularity, indicative of the 
van-Hove singularity in the density of states.
 
The plot is manifestly chiral with weighting greater for $k>0$ than
for $k<0$.  This is to be expected as we are plotting the excitations
linearized about the Fermi momentum, $+k_{Fi}$.  The heavier weighting
for $k>0$ indicates that is easier to create excitations above the Fermi
sea than below it.
It is interesting indeed that excitations below
the Fermi surface can be created at all and is a mark that interactions
are at play.

As \ref{eIIIxv} indicates, we have computed 
the single particle spectral
function  
in the neighborhood of
the Fermi points.
However in \cite{lin}, the authors found two particle spectral weight
to exist away from the Fermi
points at  harmonics $k = \pm 3k_{Fi}$ and $\om = 2m$, that is
the same energy threshold as at the Fermi points, $k_{Fi}$, themselves.
They identified the spectral weight at these points 
through combining the $SO(8)$ picture of the ladders
with the original lattice formulation.
As such, we are unable to say anything
definitive about spectral weight at these points as we are forced to
remain with the $SO(8)$ description.

The effect of a perturbation breaking $SO(8)$ can be sketched in 
a similar fashion to that of the optical conductivity.  
In the case of the single particle
contribution to the spectral function, we expect that under degenerate
perturbation theory the single parabolic line in Figure 3 to split akin
to the behaviour of the excitonic peak in the 
optical conductivity.  Again this effect will be masked by finite
temperatures which thermally broaden such features.  

At the two particle level, both in the unperturbed and perturbed cases,
several sets of two-particle pairs contribute.  In the unperturbed
case, these sets all begin to contribute at $\omega = \sqrt{4m^2 + k^2}$.
In the perturbed case, the sets will begin to contribute at 
different points.  For small shifts, a plot such as that in Figure 3
will remain much the same.  However it is worthwhile to point out 
that with each two-particle pair is associated a square-root divergence due
to the van-Hove singularity in the density of states.  Thus we would
expect to find a series of such singularities in the presence of a 
perturbation.  

\subsection{Tunneling Current}

In this section we study the tunneling between a metallic 
lead and the carbon nanotube/Hubbard ladder through a point 
contact.  Our starting point is a Lagrangian
describing the nanotube/ladder, the metallic lead, and the
tunneling interaction:
\begin{equation}\label{eIIIxvi}
\CL = \CL_{SO(8)} + \CL_{lead} + \CL_{tun} .
\end{equation}
$\CL_{SO(8)}$ is the Lagrangian of the $SO(8)$ Gross-Neveu model.

The electron gas in the lead is, in general, three dimensional.
However, in the context of tunneling through a
point contact, the electron gas can be mapped onto an one dimensional
chiral fermion (see for example
\cite{chfr,lud,ludd}).
The general idea is well illustrated by its application to the
Kondo problem.  There an electron scatters off a spin impurity at $x=0$.
The scattering is determined by the electron operator, $\psi (x=0)$.
As $\psi (0)$ only depends on its spherically
symmetric, $L=0$, mode, one can consider the scattering electron in terms of
an ingoing and outgoing radial model defined on the half-line,
$r\in [0,\infty ]$.  Unfolding the system onto the full line leaves one with
a chiral fermion.  We emphasize however that the map requires no
special symmetry; the result is exact regardless of particular anisotropies
\cite{ludd}.
As a consequence, we write $\CL_{lead}$ as
\begin{equation}\label{eIIIxvii}
\CL_{lead} = {1\over 8\pi} \Psi^\dagger \partial_{\bar{z}} \Psi ,
\end{equation}
where $\Psi$ is a massless, left moving fermion, and $z=(\tau+ix)/2$.

\newcommand{\pla}{\psi_-^\alpha (\tau )}
\newcommand{\plab}{\psi_-^{\bar{\alpha}}(\tau )}
\newcommand{\pras}{\psi_+^\alpha (\tau )}
\newcommand{\prab}{\psi_+^{\bar{\alpha}}(\tau )}
\newcommand{\ps}{\Psi (\tau )}
\newcommand{\psd}{\Psi^\dagger (\tau )}

\newcommand{\plant}{\psi_-^\alpha}
\newcommand{\plabnt}{\psi_-^{\bar{\alpha}}}
\newcommand{\prant}{\psi_+^\alpha}
\newcommand{\prabnt}{\psi_+^{\bar{\alpha}}}
\newcommand{\psnt}{\Psi}
\newcommand{\psdnt}{\Psi^\dagger}

It remains to specify $\CL_{tun}$.  In order to preserve charge, the
electrons must couple to the kinks of the $SO(8)$ 
Gross-Neveu model, the excitations
with the quantum numbers of the electron.  Thus
\begin{eqnarray}\label{eIIIxviii}
\CL_{tun} &=& ~ g_L
\bigl[
 \psd \pla + \plab\ps 
\bigr ]
 \delta (x) \cr
&& + g_R 
\bigl [
\psd\pras + \prab\ps 
\bigr ]
 \delta (x) .\end{eqnarray}
Here we have coupled the lead electrons to both the right and left moving
fields creating the kink, $\alpha$, and have allowed the two couplings,
$g_L$ and $g_R$, to be unequal.  However as we will work  
to lowest non-vanishing order in the tunneling matrix elements $g_{L/R}$,
the tunneling current will depend upon the sum,
$g_L^2 + g_R^2$, that is, the contribution of the left and right channels
to tunneling will add linearly.  Similarly, permitting other kinks to
couple to the lead electrons will give lowest order
contributions which simply add.

\vskip .75in 
\hskip -.4in
\centerline{\psfig{figure=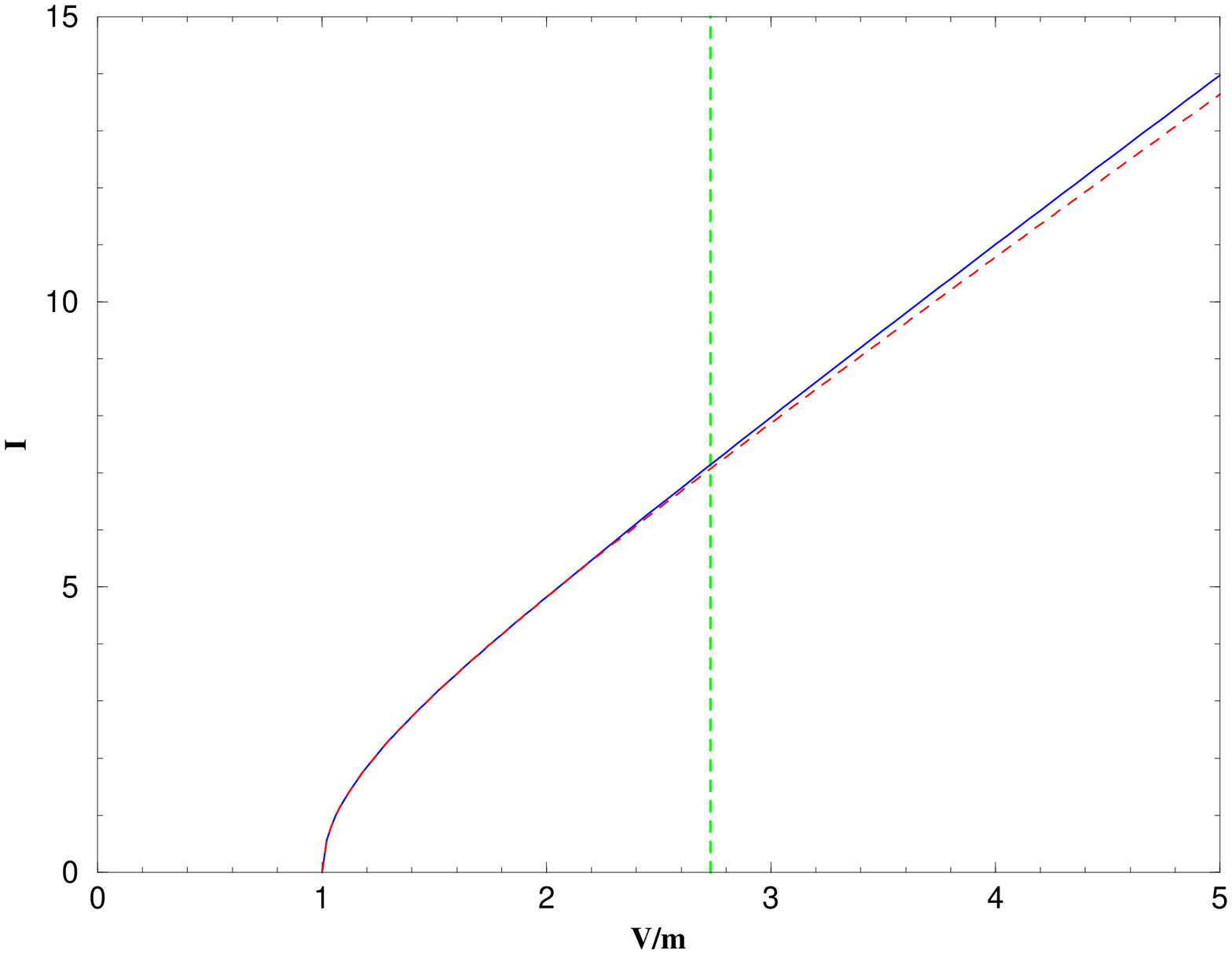,angle=-90,height=2.5in,width=2.4in}}
\vskip .25in
\indent\vbox{\noindent\hsize 3in Figure 4: Plot of the tunneling current
as a function of applied voltage.  The dashed curve describes
tunneling into a non-interacting fermionic system of mass, m.  The vertical
dashed line marks where a second set of two particle states begins 
to make a contribution.}
\vskip .25in

The calculation of the current to lowest non-vanishing order in $g_{L/R}$
follows the standard route.
The tunneling current operator is given by
\begin{eqnarray}\label{eIIIxix}
I(\tau ) &=& i g_R(\prab\ps - \psd\pras) \cr\cr
&& + i g_L(\plab\ps - \psd\pla) .
\end{eqnarray}
In order to induce current flow, $\lb I(\tau) \rb$, through the point
contact, one biases the lead with a voltage, V.
This bias can be taken into account via
a gauge transformation,
\begin{equation}\label{eIIIxx}
\ps \rightarrow e^{iV\tau}\ps .
\end{equation}
In effect we have shifted the energy levels of the electrons.
Treating the couplings, $g_L/g_R$, with linear response theory, we find
\begin{eqnarray}\nonumber
\lb I(\om )\rb &=& g_L^2 {\rm Re} \bigg\{ \int d\tau i e^{i\om \tau}
\big[\cr\cr
&& ~e^{-iV\tau}\lb \pla \plabnt (0)\rb\lb\psd\psnt (0)\rb \cr\cr
&& -e^{iV\tau}\lb \plab \plant (0)\rb\lb\ps\psdnt (0)\rb
\big]|_{{\om \rightarrow -i\om + \delta} 
\atop {V \rightarrow - iV}} \bigg\}\cr\cr
&& ~~+ (L\rightarrow R, \plant /\plabnt 
\rightarrow \prant /\prabnt).
\end{eqnarray}
\vskip -.3in\begin{equation}\label{eIIIxxi}\end{equation}
The lead electron correlator $\lb \psd \psnt (0)\rb$ is well known:
\begin{equation}\label{eIIIxxii}
\lb \psdnt (\tau ,x) \psnt (0)\rb = \lb \psnt (\tau ,x) \psdnt (0)\rb = 
{1\over \tau +ix}.
\end{equation}
With this it is straightforward to express the dc current, 
$\lb I(\om = 0)\rb$,
in terms of the single particle kink spectral function,
\begin{eqnarray}\label{eIIIxxiii}
\lb I(\om = 0)\rb &=& {1 \over 2\pi} \int^V_{-V} d \omega
\int^\infty_{-\infty} dk
\bigl [ g_L^2 A_-(\om ,k)\cr
&& \hskip 1.1in  + g_R^2 A_+(\om ,k)
\bigr ],
\end{eqnarray}
where 
$A_\pm(\om ,k) = 
A_p(\om,\pm k_{Fi}+k,k_{yi}) + A_h(-\om,\mp k_{Fi}-k,k_{yi})$, 
and $A_{p/h}$ are
the spectral functions given in \ref{eIIIxv}. 
We note that
as a technical point, in deriving the above equation we have displaced,
$\Psi$, the lead electron operator, slightly from $x=0$.  In this way
we cure the UV divergence attendant as $\tau\rightarrow 0$.  At the end
of the calculation we then take $x$ to 0.

In the previous section we have computed $A_\pm(\om ,k)$ exactly for 
energies $\om < (\sqrt{3} + 1)m$.
Inserting \ref{eIIxv} into \ref{eIIIxxiii}, we find $\lb I (0)\rb$ takes
the form,
\begin{eqnarray}\nonumber
\lb I(0) \rb &=& { |c_\pm |^2 \over m} (g_R^2 + g_L^2) (V^2 - m^2)^{1/2}
\theta (|V| -m) {\rm sgn} (V) \cr\cr
&& \hskip -.25in + \theta (|V| - 2m){\rm sgn} (V) \times 
{\rm two~particle~contribution}.
\end{eqnarray}
\vskip -.3in\begin{equation}\label{eIIIxxiv}\end{equation}
We see that for $|V| < 2m$, the system behaves as a gapped free fermion.
The first sign that there is any interaction comes for $|V| > 2m$ where
the voltage begins to probe the two particle states, a signature of
interacting fermions.

We explicitly plot $\lb I(0)\rb$ in
Figure 4.  The square root behavior near $V/m = 1$ and subsequent linear
form is typical of a gapped fermion.  At $V/m = 2$, the two particle
states begin to contribute leading to a small
change in the slope of the $I-V$
curve.  At $V/m = \sqrt{3} + 1$ (marked by the the vertical dashed line),
a second set of two 
particle states (a mass $\sqrt{3}m$ bound state together with a kink)
begin to contribute and at this point the result
ceases to be exact.  However as with the current correlators, we expect
this higher energy contribution to be small.

\newcommand{\djv}{\partial_V \lb I(0) \rb}

The change in slope in the $I-V$ curve at $V/m = 2$ can be 
explicitly computed.
To do so we consider $\djv$.  This quantity is given by
\begin{equation}\label{eIIIxxv}
\djv = {1\over \pi} (g_R^2 + g_L^2)\int^\infty_{-\infty} dk ~A_\pm(V,k) .
\end{equation}
We can thus see $\djv$ directly measures the local density of states at
$x=0$ of the nanotube/ladder system.

We plot $\djv$ in Figure 5.  The square root singularity at $V=m$ signals
the singularity of the density of states in an one dimensional system.
At $V=2m$ we see a sudden jump, indicative of the onset of the two
particle contribution.  The height of the jump can be determined
exactly:
\begin{eqnarray}\label{eIIIxxvi}
&& \djv (V/m=2^+) - \djv (V/m=2^-) = \cr\cr
&& ~~~ {8 A_F^2 \over 9 m} (g_R^2 + g_L^2)
\exp \big[ \int^\infty_0 {dx \over x}{G_f(x) \over s(x)} (1-c(x))\big].
\end{eqnarray}
The region $m<V<2m$ of $\djv$ completely determines
$m$ (by the location of the jump),
as well as  an overall scale (the product of
$(g_L^2+g_R^2)$ and the constant $A_F$, normalizing the spectral
function). 
Dividing out these non-universal numbers leaves a universal
number, characterizing the magnitude of the jump:
\begin{equation}\label{eIIIxxvii}
{8\over 9}\exp[\int^\infty_0 {dx \over x} {G_f(x) \over s(x)} (1-c(x))].
\end{equation}
This number represents a definite prediction based upon
the integrability of the model.

We now consider the approximate effect of perturbations
breaking integrability on the tunneling conductance.  As 
the tunneling conductance is determined directly from 
the single particle spectral function, we can deduce how the 
former is affected
from how the latter is changed.  At a given momentum, the single particle 
contribution to the single particle spectral function under perturbation 
comes at a discrete set of energies.  In terms of the tunneling conductance,
we expect a series of closely spaced square root divergences (a sawtooth
behaviour) about $V=m$ indicative of a series of van-Hove
singularities.  As the perturbation is removed these singularities would 
collapse on top of one another leaving the original picture in Figure 5.

In the unperturbed case the two-particle threshold is characterized
by a jump in the differential conductance.  Under a perturbation,
this jump would become a staircase or a series of smaller,
closely spaced jumps.  This is a reflection of the series of
van-Hove singularities found about $\omega = \sqrt{k^2 + 4m^2}$
in the two particle contribution to the single particle spectral function.

\vskip .75in 
\hskip -.4in
\centerline{\psfig{figure=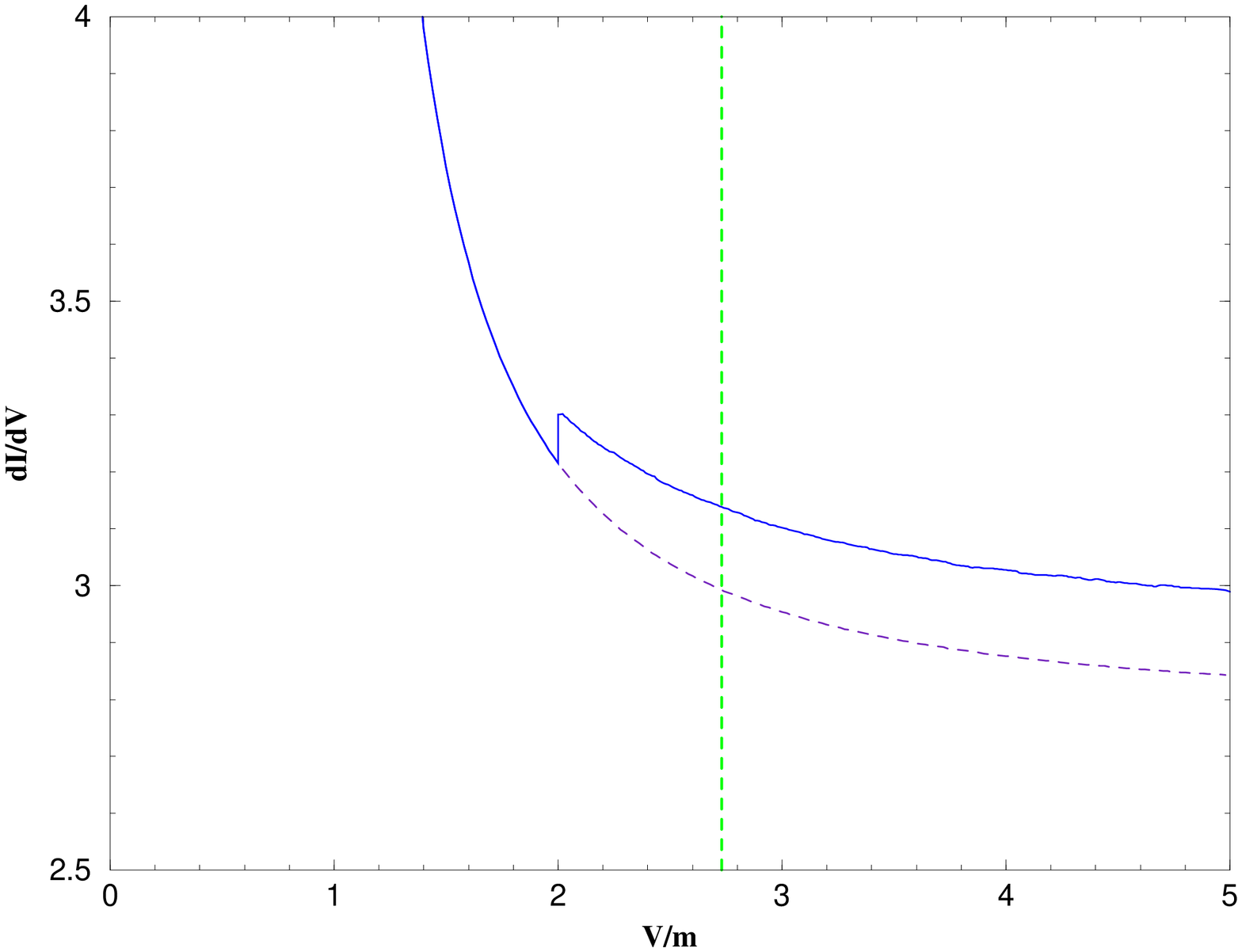,angle=-90,height=2.5in,width=2.4in}}
\vskip .25in
\indent\vbox{\noindent\hsize 3in Figure 5: Plot of the differential
conductance as a function of applied voltage.  The dashed curve
marks the single particle contribution to this quantity while the
solid curve gives both the single and first two particle contribution.
The latter plot is exact up to
$V/m = \sqrt{3} + 1$ (indicated by the dashed vertical line) where a 
bound state-kink pair begins to make a contribution.}

\section{S-matrices for $SO(8)$ Gross-Neveu}

\newcommand{\so}{\sigma_1}
\newcommand{\st}{\sigma_2}
\newcommand{\sth}{\sigma_3}
\newcommand{\ith}{i\theta/2\pi}
\newcommand{\sos}{$SO(8)$ }
\newcommand{\son}{$SO(8)$}

In this section we review the S-matrices for scattering among
the kinks and elementary (Majorana) fermions.  From these we derive
S-matrices for scattering between Dirac fermions, kinks and Dirac fermions,
and rank 2 bound states and Dirac fermions.
Knowledge of these 
S-matrices will allow us to compute the various form factors
needed to obtain the results in the previous section.
Before beginning it is necessary to lay out the group theory of $SO(8)$.
In particular it is necessary to describe the spinor representations (i.e.
the kinks) of $SO(8)$.

\subsection{$SO(8)$ Spinor Representations}

The spinor representations are expressed in terms of \sos $\gamma$-matrices.
The $\gamma$-matrices in turn are built out of the two-dimensional Pauli
matrices, $\sigma^i$'s.  As we are interested in \son, we consider four
copies of the $\sigma^i$'s,
\begin{equation}\label{eIVi}
\sigma^i_1, \sigma^i_2, \sigma^i_3, \sigma^i_4,
\end{equation}
each acting on a different two-dimensional space:
\begin{equation}\label{eIVii}
\sigma^i_n |\alpha_1, \cdots ,\alpha_4\rangle =
|\alpha_1, \cdots , \sigma^i_n \alpha_n, \cdots ,\alpha_4\rangle  .
\end{equation}
A particular basis vector, $|\alpha\rangle$,
in the corresponding 16 dimensional vector space
can be labeled by a series of four $\pm 1/2$, i.e.
\begin{equation}\label{eIViii}
\alpha = |\pm 1/2, \pm 1/2, \pm 1/2, \pm 1/2\rangle .
\end{equation}
Physically, the kink associated with $|\alpha\rangle$ then 
carries $\pm 1/2$ of
the four U(1) quantum numbers, $N_i$, corresponding to the Cartan 
elements of $SO(8)$, discussed above.

In terms of the $\sigma^i$'s, the $\gamma$-matrices are defined
by (following the conventions of
\cite{thun}),
\begin{eqnarray}\label{eIViv}
\gamma_{2n-1} &=& \sigma^1_n \otimes \prod^{n-1}_{j=1} \sigma^3_k ;\cr\cr
\gamma_{2n} &=& \sigma^2_n \otimes \prod^{n-1}_{j=1} \sigma^3_k ;
\end{eqnarray}
for $1 \leq n \leq 4$.  These matrices satisfy the necessary Clifford
algebra:
\begin{equation}\label{eIVv}
\{\gamma_a, \gamma_b\} = 2 \delta_{ab}.
\end{equation}
In this representation the Clifford-algebra generators,
$\gamma_n$, are imaginary and anti-symmetric for n even, 
while for n odd, they are real and symmetric. The $SO(8)$-generators
are represented by
\begin{equation}\label{eIVvi}
\sigma^{ab} = {1\over 2}(\gamma^a\gamma^b - \gamma^b\gamma^a),
\end{equation}
in analogy with the more familiar $SO(4)$ case.

The 16-dimensional space of the $\gamma$'s decomposes
into two 8-dimensional
spaces, each of which forms one of the two
irreducible $SO(8)$ spinor representations.
The decomposition is achieved explicitly via the hermitian 
chirality operator, 
$\Gamma$,
\begin{equation}\label{eIVvii}
\Gamma = \prod^{8}_{a=1} \gamma^a.
\end{equation}
$\Gamma$ is such that it commutes with all \sos generators and is diagonal
with eigenvalues $\pm 1$.  If $|\alpha^\pm\rangle$
is a state with an even (odd)
number of negative components (i.e. states with positive or negative
isotopic chirality), then
\begin{equation}\label{eIVviii}
\Gamma|\alpha^\pm\rangle  = \pm|\alpha^\pm\rangle .
\end{equation}
Thus the operators, $(1\pm\Gamma )/2$, project onto the two irreducible
subspaces.

The last item to be presented in this section is
the charge conjugation matrix, $C$.  In terms of the $\gamma$'s, $C$
is given by
\begin{equation}\label{eIVix}
C = \gamma_2\gamma_4\gamma_6\gamma_8 .
\end{equation}
$C$ is completely off-diagonal (as expected).  If $|\bar\alpha\rangle$ is
the anti-particle of a kink, $|\alpha\rangle$, then
\begin{equation}\label{eIVx}
|\bar\alpha\rangle = C_{\alpha\beta}|\beta\rangle.
\end{equation}
$C$ is such that $C^2 = 1$, $C^T = C$, and $C\gamma^aC^{-1} = {\gamma^a}^*$.

\subsection{S-Matrices for the Elementary Fermions and Kinks}

\newcommand{\Aa}{A_a (\theta )}
\newcommand{\Aal}{A_\alpha (\theta )}
\newcommand{\Ab}{A_b (\theta )}
\newcommand{\Abe}{A_\beta (\theta )}
\newcommand{\AB}{A_B (\theta )}

To describe scattering we introduce Faddeev-Zamolodchikov operators,
$\Aa$ and $\Aal$, that create the elementary fermions and kinks
respectively.  $\th$ is the rapidity which parameterizes a particle's
energy and momentum:
\begin{equation}\label{eIVxi}
p = m \sinh (\th ) ; ~~~~~~~ E = m \cosh (\th ) .
\end{equation}
Because the \sos Gross-Neveu model is integrable, 
scattering is completely encoded
in the two-body S-matrix.  This S-matrix, in turn, is encoded in the
commutation relations of the Faddeev-Zamolodchikov operators:
\begin{equation}\label{eIVxii}
A_1 (\th_1 ) A_2 (\th_2 ) = S^{34}_{12}(\th_{12}) 
A_4 (\th_2 ) A_3 (\th_1 ).
\end{equation}
$S^{34}_{12}(\th_{12})$ is the amplitude of a process
by which particles 1 and 2
(be they kinks,  fundamental fermions, or bosonic bound states) 
scatter into 3 and 4.  It is a function
of $\th_{12} \equiv \th_1 - \th_2$ by reason of Lorentz invariance.

Scattering between the fundamental Majorana fermions,
first determined in
\cite{zz}, is given by
\begin{equation}\label{eIVxiii}
S_{ab}^{cd} (\th ) = \delta_{ab}\delta_{cd} \so (\th ) +
\delta_{ac}\delta_{bd} \st (\th ) + \delta_{ad}\delta_{bc} \sth (\th ),
\end{equation}
where the $\sigma$'s are defined by
\begin{eqnarray}\label{eIVxiv}
\so (\th ) &=& -{i \over 3} {\pi \over i\pi - \th} \st (\th );\cr\cr
\sth (\th ) &=&  -i {\pi \over 3 \th} \st (\th );\cr\cr
\st (\th ) &=&  -Q(\th ) Q(i\pi - \th ) { s(\th ) + i \sqrt{3}/2 \over
s(\th ) - i \sqrt{3}/2} . 
\end{eqnarray}
where $s (\th ) \equiv \sinh (\th )$.  $Q$ is given to be
\begin{equation}\label{eIVxv}
Q(\th ) = {\Gamma (1/6 - \ith ) \Gamma(1/2 - \ith )
\over \Gamma (- \ith ) \Gamma(2/3 - \ith ) }.
\end{equation}
$S_{ab}^{cd}(\th )$ satisfies a Yang-Baxter equation,
\begin{eqnarray}\label{eIVxvi}
S^{c_1c_2}_{a_1a_2}(\th_{12}) && S^{b_1c_3}_{c_1a_3}(\th_{13})
S^{b_2b_3}_{c_2c_3}(\th_{23}) = \cr\cr
&& S^{c_2c_3}_{a_2a_3}(\th_{23})
S^{c_1b_3}_{a_1c_3}(\th_{13}) S^{b_1b_2}_{c_1c_2}(\th_{12}).
\end{eqnarray}
Physically, the Yang-Baxter equation encodes the equivalence of
different ways of representing three-body interactions in terms of
two-body amplitudes.  Formally, it expresses the associativity of
the Faddeev-Zamolodchikov algebra.
The S-matrix, $S_{ab}^{cd}(\th )$, also satisfies both a crossing,
\begin{equation}\label{eIVxvii}
S^{cd}_{ab}(\th ) = S^{da}_{bc}(i\pi - \th ),
\end{equation}
and unitarity relation,
\begin{equation}\label{eIVxviii}
S^{cd}_{ab}(\th )S^{ef}_{cd}(-\th ) = \delta_a^e\delta_b^f.
\end{equation}
These constraints determine $S^{cd}_{ab}$ up to a `CDD'-factor.
Such factors determine the pole structure in the physical strip,
${\rm Re} (\th ) = 0$,
$0 < {\rm Im}{\th} < 2\pi$, of the scattering matrix
and so are indicative of bound states.  Here the CDD factor is
\begin{equation}\label{eIVxix}
{s(\th ) + i \sqrt{3}/2 \over s(\th ) - i \sqrt{3}/2 },
\end{equation}
and reflects the fact that two fermions can form a bound state of
mass $\sqrt{3}m$ transforming
as a component of the 29 dimensional representation of the $SO(8)$ Yangian.
As stated in the introduction, this representation is formed out of
a rank-2 anti-symmetric $SO(8)$ tensor and a scalar.
The overall sign of the S-matrix is determined by examining the residue
of the pole in $S^{aa}_{aa}$ at $\th = i\pi/3$.  This pole is
indicative of the formation of a mass $\sqrt{3}m$ scalar bound state in the
s-channel and so should have positive imaginary residue.

Scattering between the kinks and fundamental fermions
is given by
\begin{equation}\label{eIVxx}
S_{a\alpha}^{b\beta} = -t_1(\th ) \delta_{ab} \delta_{\alpha\beta}
- t_2 (\th ) \sigma^{ba}_{\beta\alpha}.
\end{equation}
The form of \ref{eIVxviii}, first written down in
\cite{witt},
is determined by writing
down all possible $SO(8)$ invariant tensors with indices $a,b$ and $\alpha,
\beta$.  
$t_1$ and $t_2$ are described by
\begin{eqnarray}\label{eIVxxi}
t_1 (\th ) - t_2 (\th ) &=& {\th - i2\pi/3 \over \th + i2\pi/3} 
S_2 (\th );\cr\cr
t_1 (\th ) + 7 t_2 (\th ) &=& S_2 (\th) ,
\end{eqnarray}
where $S_2(\th )$ is given by
\begin{equation}\label{eIVxxii}
S_2 (\th ) = { s(\th ) + i \sqrt{3}/2 \over s(\th ) - i \sqrt{3}/2}
{\Gamma (5/6 + \ith ) \Gamma (4/3 - \ith) \over \Gamma (5/6 - \ith)
\Gamma (4/3 + \ith )}. 
\end{equation}
Again, the Yang-Baxter equation, crossing, and unitarity are 
used to determine
\ref{eIVxxi} and \ref{eIVxxii}.  These constraints were first 
explicitly solved in \cite{thun}.
The overall sign is fixed by insisting the pole in $S_{a\alpha}^{a\alpha}$
at $\th = i2\pi/3$, indicative of a kink formed in the s-channel
as a fermion-kink bound state, has positive imaginary residue.

Kink-kink scattering takes the form
\begin{equation}\label{eIVxxiii}
S_{\alpha\beta}^{\gamma\delta} = {1 \over 16} \sum^{8}_{r=0}
{u_r (\th ) \over r!}
\sigma^{(r)}_{\gamma\beta} \sigma^{(r)}_{\delta\alpha},
\end{equation}
where $\sigma^{(r)}$ is a rank-r anti-symmetric tensor:
\begin{equation}\label{eIVxxiv}
\sigma^{(r)} \equiv \sigma^{a_1 \cdots a_r}
= (\gamma^{a_1} \cdots \gamma^{a_r})_A .
\end{equation}
Here A represents a complete antisymmetrization of the gamma matrices.
By $\sigma^{(r)}_{\gamma\beta} \sigma^{(r)}_{\delta\alpha}$ we mean a
trace over all possible rank-r anti-symmetric tensors:
\begin{equation}\label{eIVxxv}
\sigma^{(r)}_{\gamma\beta} \sigma^{(r)}_{\delta\alpha} =
\sum_{a_1 \cdots a_r} \sigma^{a_1 \cdots a_r}_{\gamma\beta}
\sigma^{a_1 \cdots a_r}_{\delta\alpha}.
\end{equation}
Again the generic form of $S_{\alpha\beta}^{\gamma\delta}$ was determined
in \cite{witt}, 
while the specific forms of the $u$'s were given in \cite{thun}.
There it was found
\begin{eqnarray}\label{eIVxxvi}
u_{4+r} (\th ) &=& (-1)^r u_{4-r}(\th ) ;\cr\cr
u_{r+2} (\th ) &=& u_r (\th ) { (1 - r/3) - (1 + i\th/\pi ) \over
(1 - r/3) + (1 + i \th/\pi )} ;\cr\cr
u_2 (\th ) &=& \st (i\pi - \th ) - \sth (i\pi - \th ) ;\cr\cr
u_1 (\th ) &=& S_2 (i\pi - \th).
\end{eqnarray}
The $SO(8)$ Gross-Neveu 
model has an isotopic chirality conservation law \cite{thun}.
Thus opposite chirality kink scattering is determined solely 
by $u_n$, $n$ odd,
while same chirality kink scattering is
determined solely by $u_n$, $n$ even.  The overall sign of the 
S-matrix
again is chosen so that appropriate s-channel poles
have positive imaginary residues.  That the $u$'s bear a close
relationship to amplitudes from fermion-fermion scattering and kink-fermion
scattering is a reflection of the triality symmetry of $SO(8)$.

\subsection{S-Matrices for Dirac Fermions}

\newcommand{\dpm}{D^\pm_i(\th )}
\newcommand{\dpo}{D^+_i(\th )}
\newcommand{\dm}{D^-_i(\th )}

On occasion the Majorana fermion basis will not be the most convenient.
For example, 
as these fermions do not carry definite U(1) charge, they will not couple
nicely to a chemical potential introduced when the system is doped.
Thus we would want to consider, as in \cite{doped}, a
fermion basis with well-defined U(1) charges, 
i.e. we would want to reexpress
the Majorana fermions as Dirac-fermions.

As there are eight Majorana fermions in $SO(8)$ Gross-Neveu, we have four
Dirac fermions.  Denoting the corresponding Faddeev-Zamolodchikov operators
by $\dpm$, where $\dm$ is $\dpo$'s anti-particle, we
have
\begin{equation}\label{eIVxxvii}
\dpm = {1\over \sqrt{2}} (A_{2i}(\th ) \pm i A_{2i-1}(\th )) ~~~
1\leq i \leq 4 ,
\end{equation}
where the $A_i$'s are Majorana fermions.
Because scattering is determined by the commutation relations,
\begin{equation}\label{eIVxxviii}
D^{c_i}_i (\th_1) D^{c_j}_j(\th_2) = S^{(kc_k)(lc_l)}_{(ic_i)(jc_j)}
D^{c_k}_k(\th_2) D^{c_l}_l(\th_1) ,
\end{equation}
where $c = \pm 1$ is indicative of the U(1) charge carried,
we can determine $S^{(kc_k)(lc_l)}_{(ic_i)(jc_j)}$ from \ref{eIVxiii}.
We find the following,
\begin{eqnarray}\label{eIVxxix}
S^{(kc_k)(lc_l)}_{(ic_i)(jc_j)} (\th ) &=& \delta_{ij}\delta_{kl}
C^{c_ic_j}_D C^{c_kc_l}_D \so (\th ) \cr\cr
&& + \delta_{ik}\delta_{c_ic_k}\delta_{jl}\delta_{c_lc_j}\st (\th ) \cr\cr
&& + \delta_{il}\delta_{c_ic_l}\delta_{jk}\delta_{c_kc_j}\sth (\th ) ,
\end{eqnarray}
where $C_D \equiv \sigma_1$ is the charge conjugation matrix for
Dirac fermions.  We point out that two Dirac fermions of the 
same U(1) charge
will scatter diagonally, i.e.
\begin{equation}\label{eIVxxx}
D^\pm_i(\th_1 ) D^\pm_i(\th_2 ) = (\st (\th_{12}) + \sth (\th_{12}))
D^\pm_i(\th_2 ) D^\pm_i(\th_1 ) .
\end{equation}

We will also want to consider scattering between kinks and Dirac fermions.
From \ref{eIVxx} we find
\begin{eqnarray}\label{eIVxxxi}
S^{(jc_j)\beta}_{(ic_i)\alpha} (\th ) &=&
\delta_{ij}\delta_{c_ic_j}\delta_{\alpha\beta}(t_1(\th ) - t_2(\th ))\cr\cr
&& ~~~~~~~
+ {c_ic_j \over 2}(\Gamma^j_{c_j}\Gamma^i_{-c_i})_{\beta\alpha}t_2 (\th ) .
\end{eqnarray}
where
\begin{equation}\label{eIVxxxii}
\Gamma^j_{\pm} = \gamma^{2j-1} \pm i\gamma^{2j} .
\end{equation}
When $\Aal$ carries U(1) charge $c_i/2$ (i.e.  when the U(1) charges
of $\Aal$ and $D_i$ carry the same sign) the above form simplifies greatly:
\begin{equation}\label{eIVxxxiii}
S^{(ic_i)\beta}_{(ic_i)\alpha} (\th ) =
t_1(\th ) - t_2(\th ) ,
\end{equation}
i.e. scattering becomes diagonal.  This is exploited
in \cite{doped} to compute the excitation
energy of a kink in the doped system.

We want to consider one last S-matrix involving the Dirac fermions,
that of a Dirac fermion with rank 2 tensorial bound states.  We will not
do this in general.  Rather we consider a bound state carrying two
non-zero positive charges, say $U_i = +1$ and $U_j = +1$, scattering off a
Dirac fermion carrying charge either $U_i = +1$ or $U_j = +1$.  This is
precisely the situation encountered when treating spin 1 excitations
in the doped system.

We thus introduce the bound state Faddeev-Zamolodchikov operator, $\AB$.
$\AB$ can be thought of as a bound state of two Dirac fermions with charge
$U_i=+1$ and $U_j=+1$, and consequently can be represented in terms of
the corresponding
two Faddeev-Zamolodchikov operators,
\begin{eqnarray}\label{eIVxxxiv}
i g^B_{(i+)(j+)} \AB &=& \cr\cr
&& \hskip -1.in {\rm res}_{\delta = 0}~
A_{i+}(\th + \delta + i\bar{u}^{(j-)}_{(i+)\bar{B}})
A_{j+}(\th - i\bar{u}^{(i-)}_{(j+)\bar{B}}), 
\end{eqnarray}
where ${\rm res}_{\delta =0}$ denotes the residue at $\delta = 0$ and
$\bar{B}$ is the charge conjugate of $B$.  In writing \ref{eIVxxxiv},
we have taken the particle normalization to be
$\langle \th | \th '\rangle = 2\pi \delta (\th - \th ')$.
The $\bar{u}$'s are given
by
\begin{equation}\label{eIVxxxv}
\bar{u}^{s'}_{sB} = \pi - u^{s'}_{sB} ,
\end{equation}
where $iu^{s'}_{sb}$ is the location of the pole indicative of the
particle $s'$ in the
$s\negthinspace-\negthinspace\negthinspace B$
S-matrix.  Here we have
\begin{equation}\label{eIVxxxvi}
u^{(j-)}_{(i+)\bar{B}} = u^{(i-)}_{(j+)\bar{B}} = 5\pi /6 .
\end{equation}
The $g$'s are related to the residues of the poles in the
$(i+)\negthinspace\negthinspace-\negthinspace\negthinspace(j+)$
S-matrix and can be interpreted as the amplitude to form
the bound state from $(i+)$ and $(j+)$.  If $u^B_{(i+)(j+)}$ is the location
of the pole indicative of $B$, $g^B_{(i+)(j+)}$ is defined by
\begin{equation}\label{eIVxxxvii}
S^{(i+)(j+)}_{(i+)(j+)} (\th ) \sim
i {g^B_{(i+)(j+)} g_B^{(i+)(j+)}
\over \theta - i u^B_{(i+)(j+)}} .
\end{equation}
It is then easy to show (from \ref{eIVxxx}) that
\begin{eqnarray}\label{eIVxxxviii}
g^B_{(i+)(j+)} g^{(i+)(j+)}_B &=& S_0 ;\cr
g^B_{(j+)(i+)} g^{(i+)(j+)}_B &=& -S_0 ;
\end{eqnarray}
where $S_0$ is some constant.  Hence $g^B_{(j+)(i+)} /g_{(i+)(j+)}^B = -1$.
This last relation is all we will need to determine the scattering between
$\AB$ and $D^+_{i,j}$.

From \ref{eIVxxxii} and \ref{eIVxxviii} we find that
$S^{B(i/j+)}_{B(i/j+)}$ is given by
\begin{eqnarray}\nonumber
S^{B(i/j+)}_{B(i/j+)}(\th ) &=& \st (\th - i\pi/6)
(\st (\th + i\pi/6)\! +\! \sth (\th + i\pi/6) ) \cr\cr
&& ~~~~~~~~~~~~ - \sth (\th - i\pi/6) \sth (\th + i\pi/6) \cr\cr
&=& {\th + i{\pi\over 6} \over \th - i{\pi\over 6}}
\big(\sigma_2 (\th +i\pi/6)
+ \sigma_3 (\th +i\pi/6)\big) \cr\cr
&& ~~~~~~\times\big(\sigma_2 (\th -i\pi/6) + \sigma_3(\th -i\pi/6)\big), 
\end{eqnarray}
\vskip -.5in
\begin{equation}\label{eIVxxxix}\end{equation}
so that scattering between $\AB$ and $D^+_{i,j}$ is diagonal.

\section{Form Factors}

Here we determine the needed form factors to compute
correlators in the undoped system.  If the reader is uninterested in
the actual derivation, the results are summarized at
the end of the section.

\subsection{Basic Properties: Two Particle Form Factors}

\newcommand{\fp}{f^\psi}

The two particle form factors of a field $\psi (x)$ are defined as the
matrix elements,
\begin{equation}\label{eVi}
f^\psi_{12} (\th_1 , \th_2 ) = \langle \psi (0) A_2(\th_2) 
A_1(\th_1) \rangle .
\end{equation}
The form of $\fp_{12}(\th_1 ,\th_2 )$ is constrained by integrability,
braiding relations, Lorentz invariance, and hermiticity.

The constraint coming from integrability
arises from the scattering of Faddeev-Zamolodchikov
operators.  For \ref{eVi} to be consistent with \ref{eIVxii}, we must have
\begin{equation}\label{eVii}
\fp_{21}(\th_2 ,\th_1 ) = S^{34}_{12}(\th_{12})\fp_{34}(\th_1,\th_2).
\end{equation}
The second constraint can be thought of as a periodicity axiom.
It reads
\begin{equation}\label{eViii}
\fp_{21}(\th_2,\th_1) = R_{\psi_2\psi}\fp_{12}(\th_1-2\pi i,\th_2).
\end{equation}
$R_{\psi_2\psi}$ is a phase\footnote{In certain 
circumstances $R$ is a matrix,
thus marking out the difference between abelian and non-abelian braiding.}
that arises from the non-locality of the fields in \sos Gross-Neveu.
This non-locality leads to non-trivial braiding relations 
between fields which
$R_{\psi_2\psi}$ encodes:
\begin{equation}\label{eViv}
\psi_2 (x,t)\psi (y,t) = R_{\psi_2\psi}\psi (y,t)\psi_2 (x,t); 
~~~~~~~x < y .
\end{equation}
$\psi_2$ here is the field that is associated with the particle,
$A_2(\th_2)$.  \ref{eViii} can be derived through crossing symmetry constraints
(see, for example, \cite{double,smir}).

The form factor, $\fp_{12}$, must also satisfy Lorentz invariance.  If
$\psi$ has Lorentz spin, $s$, $\fp_{12}$ will take the form
(at least in the cases at hand),
\begin{equation}\label{eVv}
\fp_{12}(\th_1,\th_2 ) = e^{s(\th_1+\th_2)/2}f(\th_{12}),
\end{equation}
where by virtue of $\th_{12} \equiv \th_1 - \th_2$, $f$ is a Lorentz
scalar.

The constraints \ref{eVii}, \ref{eViii}, and \ref{eVv}, do not 
uniquely specify $\fp_{12}$.
It is easily seen that if $\fp_{12}$ satisfies these axioms then so does
$\fp_{12}R(\cosh (\th_{12}))$, where $R(x)$ is some rational expression.  
The
strategy is then as follows.  One first determines the minimal solution to
the constraints, minimal in the sense that is has the minimum number 
of zeros
and poles in the physical strip,
${\rm Re}(\th ) = 0, 0 < {\rm Im}(\th ) < 2\pi$.  Then one adds poles 
according
to the theory's bound state structure.  If the S-matrix element scattering
particles 1,2 has a pole at $\th = iu$, then $\fp_{12}(\th_1,\th_2)$ has
a pole at
\begin{equation}\label{eVvi}
\th_2 = \th_1 + iu .
\end{equation}
Insisting that $\fp_{12}$ has such poles and only such poles fixes 
$R(x)$ up to a constant.

The phase of this constant can be readily determined.  Appealing
to hermiticity gives us
\begin{eqnarray}\label{eVvii}
\langle \psi (0) A_2(\th_2) A_1(\th_1) \rangle^*  &=&
\langle A_1^\dagger(\th_1) A_2^\dagger(\th_2) \psi^\dagger (0) \rangle\cr\cr
&& \hskip -.5in= 
\langle \psi^\dagger (0) A_{\bar 1}(\th_1-i\pi) A_{\bar 2}(\th_2-i\pi)
\rangle,\end{eqnarray}
where the last line follows by crossing and so
\begin{equation}\label{eVviii}
f^\psi_{12}(\th_1 ,\th_2)^* =
f^{\psi^\dagger}_{\bar{2}\bar{1}}(\th_2-i\pi,\th_1-i\pi).
\end{equation}
For the form factors we will examine, this will be enough to fix their
overall phase.

\subsection{Basic Properties: One Particle Form Factors}

One particle form factors are in a sense trivial;  Lorentz invariance
completely determines their form.  If $\psi(x)$ has Lorentz spin, $s$, 
then
\begin{equation}\label{eVix}
\fp_1(\th) = \langle \psi (0)A_1 (\th)\rangle = c e^{s\th} ,
\end{equation}
where $c$ is some constant.  To determine $c$ we use the theory's
bound state structure.

In analogy with \ref{eIVxxxiv} we write
\begin{equation}\label{eVx}
i g^1_{23}A_1(\th ) =  {\rm res}_{\delta = 0}~
A_2(\th + \delta + i \bar{u}^{\bar{3}}_{2\bar{1}})
A_3(\th - i \bar{u}^{\bar{2}}_{3\bar{1}}) .
\end{equation}
Then we have
\begin{equation}\label{eVxi}
i g^1_{23}\fp_1(\th ) =  {\rm res}_{\delta = 0}~
\fp_{32}(\th - i\bar{u}^{\bar{2}}_{3\bar{1}},
\th + \delta + i \bar{u}^{\bar{3}}_{2\bar{1}}).
\end{equation}
Thus knowledge of the two particle form factors completely determines
their one-particle counterparts.

\subsection{Braiding of the Fields}

\newcommand{\rot}{{_P R}_{12}}
In order to employ the periodicity axiom \ref{eViii}, 
we need to specify
the braiding of the fields.  In order to do this for the kink and
fermion fields, we identify these fields with their corresponding
excitations, $A_\alpha$ and $A_a$.  The braiding of the fields is then
encoded in the asymptotic limits of the corresponding S-matrices
(see \cite{smir}).
Precisely, if ${_P R}_{12}$ is defined by (where $P=\pm$ denotes right/left
movers)
\begin{equation}\label{eVxii}
\psi_{1P}(x,t)\psi_{2P}(y,t) = {_P R}_{12}\psi_{2P}(y,t)\psi_{1P}(x,t),
~~~ x < y ,
\end{equation}
then $\rot$ is given by 
\begin{equation}\label{eVxiii}
\rot = (S^{12}_{12} (+\infty ))^P .
\end{equation}
In this way we find 
\begin{equation}\label{eVxiv}
{_\pm R}_{\alpha\beta}=\left\{ \begin{array}{ll}
-1, & \mbox{$\alpha$,$\beta$ are the same chirality}\\
 \pm i, & \mbox{$\alpha$,$\beta$ are of opposite chirality;}
\end{array}\right.
\end{equation}
and
\begin{equation}\label{eVxv}
{_\pm R}_{a\beta} = \pm i .
\end{equation}
The braiding relationship among the kinks differ from what one finds
by defining the kink fields as vertex operators,
\begin{equation}\label{eVxvi}
\psi_{\alpha P} = e^{i\alpha\cdot\bar{\phi}},
\end{equation}
and using the easily derived braiding relations:
\begin{equation}\label{eVxvii}
e^{ia\phi_P(x,t)}e^{ib\phi_P(y,t)}
= e^{iP\pi ab}e^{ib\phi_P(y,t)}e^{ia\phi_P(x,t)}, ~~~~ x>y.
\end{equation}
Similarly, the correct braiding of kinks with the Majorana fermions
is not obtained by considering the braiding of the corresponding vertex
operators.

The only other fields that we concern ourselves with in this paper are
the currents.  However the current are local fields and so have trivial
braiding with the kinks and the fermions.

\subsection{Two Particle Form Factors for the $SO(8)$ Currents}

\newcommand{\fcf}{_\mu f^{ab}_{cd}}
\newcommand{\fcft}{_\mu f^{ab}_{cd}(\th_1,\th_2)}
\newcommand{\fck}{_\mu f^{ab}_{\alpha\beta}}
\newcommand{\fckt}{_\mu f^{ab}_{\alpha\beta}(\th_1,\th_2)}
\newcommand{\cop}{G^{ab}_\mu}

There are two possible two particle form factors for the $SO(8)$ currents,
one involving two fermions,
\begin{equation}\label{eVxviii}
\fcft = \langle \cop(0) A_d(\th_2 )A_c(\th_1)\rangle ,
\end{equation}
and one involving two kinks,
\begin{equation}\label{eVxix}
\fckt = \langle \cop(0) A_\beta (\th_2 )A_\alpha (\th_1)\rangle ,
\end{equation}
where the two kinks have the same chirality.  By the triality
symmetry of \son , these two form factors will have the same functional
form.  Indeed we will use triality to set their relative normalization,
crucial to the calculation of the conductivity in Section 3.

\vskip .2in
\noindent{\bf Calculation of $\fcf$:}
\vskip .2in

To determine $\fcf$, one first fixes its group theoretical structure.
Given that $\cop$ is anti-symmetric and that $A_dA_c$ must be anti-symmetric
in $d$ and $c$ in order to couple to $\cop$ \cite{sla} , we
must look for an invariant tensor anti-symmetric in the pairs
$a,b$ and $c,d$.  The only choice is
$\delta_{ac}\delta_{bd} - \delta_{ad}\delta_{bc}$, and so $\fcf$ takes
the form,
\begin{equation}\label{eVxx}
\fcft = (\delta_{ac}\delta_{bd} - \delta_{ad}\delta_{bc}) 
f_\mu (\th_1,\th_2) .
\end{equation}
Lorentz invariance demands
\begin{equation}\label{eVxxi}
f_\mu (\th_1,\th_2) =
\bigl (
e^{(\th_1+\th_2)/2} - (-1)^\mu e^{-(\th_1+\th_2)/2}\bigr )
f(\th_{12}) .
\end{equation}
Having so constrained the form of $\fcft$, conditions \ref{eVii} and 
\ref{eViii} tell us $f(\th_{12})$ must satisfy
\begin{eqnarray}\label{eVxxii}
f(-\th ) &=& -f(\th )(\st (\th ) - \sth (\th )) ;\cr\cr
f(-\th ) &=& f(\th - 2\pi i) .\end{eqnarray}
As the current is bosonic, the braiding here is trivial.
Because $\st - \sth$ can be written in the form,
\begin{eqnarray}\label{eVxxiii}
\st (\th) - \sth (\th ) &=& \exp \bigg[ \int^\infty_0 {dx \over x}
s({x\th \over i\pi}) G_c(x) \bigg] ,\cr\cr\cr
G_c(x) &=& 2{c(x/6) - s(x/6) e^{-2x/3} \over c(x/2)},\end{eqnarray}
it is readily checked that
\begin{equation}\label{eVxxiv}
f(\th ) = s(\th /2)\exp \bigg[\int^\infty_0 {dx \over x} {G_c(x) \over s(x)}
\sin^2({x\over 2\pi}(i\pi + \th))\bigg] ,
\end{equation}
minimally satisfies the above condition.  To add the bound state structure,
we note that the two fermions $c,d$ form a bound state of mass $\sqrt{3}m$.
This implies a pole in the c-d S-matrix at $\th = iu = i\pi/3$, and so
a pole in $f(\th_{12})$ at $\th_2 = \th_1 + i\pi/3$.  Hence we multiply
$f(\th)$ by $1/(c(\th)-1/2)$, leading us to the form for $\fcft$:
\begin{eqnarray}\label{eVxxv}
\fcft &=& iA_G (\delta_{ac}\delta_{bd} - \delta_{ad}\delta_{bc})\cr\cr
&& \hskip -.6in \times (e^{(\th_1+\th_2)/2} - (-1)^\mu e^{-(\th_1+\th_2)/2})
{s(\th_{12}/2) \over c(\th_{12}) - 1/2}\cr\cr
&& \hskip -.6in \times \exp [\int^\infty_0 {dx \over x} {G_c(x) \over s(x)}
\sin^2({x\over 2\pi}(i\pi + \th_{12}))] ,
\end{eqnarray}
where $i A_G$ is some normalization with mass dimension
$[m]$.

The phase of $A_G$ is determined through the hermiticity condition
\begin{equation}\label{eVxxvi}
\fcft^* = {_\mu f}^{ab}_{dc}(\th_2 - i\pi,\th_1-i\pi) .
\end{equation}
This implies that $A_G$ is real.

\vskip .2in
\noindent{\bf Calculation of $\fck$:}
\vskip .2in
\newcommand{\cs}{(C \sigma^{ab})_{\alpha\beta}}
\newcommand{\sco}{(\sigma^{ab}C)_{\alpha\beta}}

Again we begin by identifying the group theoretical structure of
$\fck$:
\begin{equation}\label{eVxxvii}
\fckt = \cs f_\mu (\th_1,\th_2) ,
\end{equation}
where $C$ is the charge conjugation matrix introduced in Section 4.
$\cs$ is not only anti-symmetric in $a,b$ but also in $\alpha ,\beta$ as
it must be if the kinks are to couple to $\cop$ \cite{sla}.  Now $\cs$
is not the only available invariant tensor.  One also has
$\sco$, anti-symmetric combinations built up out of $\gamma^a C \gamma^b$
or $C\gamma^a C \gamma^b C$, or some combination of all three (but not
$\sigma^{ab}$ as this choice violates obvious U(1) conservation).  Perhaps
the most natural choice is to make $\fck$ explicitly C-symmetric:
\begin{equation}\label{eVxxviii}
\fckt = (\cs + \sco ) f_\mu (\th_1,\th_2).
\end{equation}
However this forces $_\mu f^{(2a-1)(2a)}_{\alpha\beta}$ to zero and this
is not consistent with triality, i.e. all $_\mu f^{(2a-1)(2a)}_{cd}$
do not vanish.  So we choose as above.  Note that we could
equally well have chosen $\sco$.  That our choice is not C-symmetric is not
surprising as the theory is not trivially C-symmetric.  With kink-fermion
scattering, a C-transformation changes the sign of selected amplitudes
(see \ref{eIVxx}).

As before Lorentz invariance demands
\begin{equation}\label{eVxxix}
f_\mu (\th_1,\th_2) =
(e^{(\th_1+\th_2)/2} - (-1)^\mu e^{-(\th_1+\th_2)/2})f(\th_{12}) .
\end{equation}
Then by \ref{eVii}, $f(\th_{12})$ must satisfy
\begin{eqnarray}\label{eVxxx}
f(-\th ) &=& -u_2 (\th) {i\pi + 3\th \over 4\pi i - 3\th}{i\pi - \th 
\over \th}
f(\th), \cr\cr
&=& -(\st (\th) - \sth (\th ))f(\th ), \end{eqnarray}
where the last line follows from \ref{eIVxxvi}. That \ref{eVxxx} has exactly
the same form as \ref{eVxxii} is a non-trivial consequence of triality.
The periodicity axiom also takes the same form as before:
\begin{equation}\label{eVxxxi}
f(-\th ) = f(\th - 2\pi i) .
\end{equation}
As the bound states of two fermions are identical to that of two
same chirality kinks,
the bound state structure of $\fck$ is the same as $\fcf$.  Hence
\begin{eqnarray}\label{eVxxxii}
\fckt &=& i B \cs \cr\cr
&& \hskip -.8in \times \bigl (
e^{(\th_1+\th_2)/2} - (-1)^\mu e^{-(\th_1+\th_2)/2}
\bigr )
{s(\th_{12}/2) \over c(\th_{12}) - 1/2}\cr\cr
&& \hskip -.8in 
\times \exp \bigg[\int^\infty_0 {dx \over x} {G_c(x) \over s(x)}
\sin^2({x\over 2\pi}(i\pi + \th_{12}))\bigg] .\end{eqnarray}
What is left is to determine the relative value of $B$ to $A_G$ 
via triality.
An overall normalization is not so interesting when computing two-point
correlators as the physics so encoded is not universal.

\newcommand{\got}{G^{12}_0}
To determine the normalization we focus upon a particular current component
$\got$.  $\got$ is given by
\begin{equation}\label{eVxxxiii}
\got \propto \del_x \phi_1 ,
\end{equation}
where $\phi_1$ is one of the four Cartan bosons.  Under the triality
transformation of \ref{eIIxxv}, $\got$ is transformed into
\begin{equation}\label{eVxxxiv}
\got \rightarrow  {1\over 2} (\got + G^{34}_0 + G^{56}_0 + G^{78}_0).
\end{equation}
Now focus upon the form factor $f^{12}_{\alpha\beta}$ where
$|\alpha \rangle = (1/2,1/2,1/2,1/2)$ and
$|\beta \rangle = (-1/2,-1/2,-1/2,-1/2)$.  Under a naive counting of
quantum numbers, $A_\alpha$ and $A_\beta$ are transformed into Dirac
fermions:
\begin{eqnarray}\label{eVxxxv}
A_\alpha &&\rightarrow {1 \over \sqrt{2}}(A_2 + i A_1) \cr
A_\beta &&\rightarrow {1 \over \sqrt{2}}(A_2 - i A_1) .
\end{eqnarray}
Hence we have
\begin{equation}\label{eVxxxvi}
f^{12}_{\alpha\beta} \rightarrow {i\over 2}f^{12}_{12} .
\end{equation}
This fixes $B = A_G/2$.

\subsection{One Particle Form Factors for $SO(8)$ Currents}

\newcommand{\acd}{A_{\{cd\}}}
\newcommand{\gcdo}{g^{\{cd\}}_{cd}}

Only the rank-2 tensorial states couple individually to the $\cop$'s.
Denoting the Faddeev-Zamolodchikov operators for the bound states by
$\acd$ we have (as in \ref{eIVxxxiv})
\begin{equation}\label{eVxxxvii}
i g^{cd}_{\{cd\}} \acd (\th) = {\rm res}_{\delta = 0}~
A_c(\th + \delta + i\pi/6)A_d(\th-i\pi/6) ,
\end{equation}
where $A_c,A_d$ are Majorana fermions.  $\gcdo$, the amplitude to form the
bound state $\acd$ from $A_c$ and $A_d$, is given by the residue in
$S^{cd}_{cd}$ at $iu^{\{cd\}}_{cd}$:
\begin{equation}\label{eVxxxviii}
S_{cd}^{cd} (\th )
\sim i {\gcdo g^{cd}_{\{cd\}} \over \th - iu^{\{cd\}}_{cd}},
\end{equation}
and so is readily computed to be
\begin{equation}\label{eVxxxix}
\gcdo = (2 \sqrt{3\pi} {\Gamma (2/3) \over \Gamma (1/6)})^{1/2}.
\end{equation}
Note that the value of $\gcdo$ given here is ambiguous up to a phase.
However the phase can be pinned down through an appeal to hermiticity.

Using \ref{eVxxxvii}, we can write down
\begin{eqnarray}\label{eVxl}
_\mu f^{ab}_{\{cd\}} (\th ) &=& i A_G
(\delta_{ac}\delta_{bd} - \delta_{ad}\delta_{bc})\cr\cr
&& \hskip -.4in \times (e^\th - (-1)^\mu e^{-\th})
{1 \over \sqrt{3}}
\bigg( 2\sqrt{3\pi} {\Gamma (2/3) \over \Gamma (1/6)}\bigg)^{-1/2}\cr\cr
&& \hskip -.4in \times
\exp \bigg[-\int^\infty_0 {dx \over x} {G_c(x) \over s(x)} 
s^2(x/3)\bigg] .\end{eqnarray}
The overall phase (up to a sign) has been fixed by the constraint,
$_\mu {f^{ab}_{\{cd\}}}^* =  - _\mu{f^{ab}_{\{cd\}}}$, 
arising from hermiticity.
Although the overall sign is uncertain, the computation
of correlators does not depend on its value.

\subsection{Two Particle Form Factors for Gross-Neveu Fermion Operators}
\newcommand{\ff}{\psi^a_\pm}
\newcommand{\fkk}{_\pm f^a_{\alpha\beta}}
\newcommand{\fkkt}{_\pm f^a_{\alpha\beta}(\th_1,\th_2)}
\newcommand{\cg}{(C\gamma^a)_{\alpha\beta}}
\newcommand{\gc}{(\gamma^aC)_{\alpha\beta}}
\newcommand{\fpm}{{f_\pm (\th_1,\th_2)}}
\newcommand{\fo}{f_0}

At the two particle level, two kinks of opposite chirality will
couple to a right/left fermion field, $\ff$.  This leads to form factors,
$\fkkt$, defined by
\begin{equation}\label{eVxli}
\fkkt = \langle \ff (0) A_\beta(\th_2)A_\alpha(\th_1)\rangle .
\end{equation}
Covariance under $SO(8)$ demands $\fkk$ takes the form
\begin{equation}\label{eVxlii}
\fkkt = [ c_1 \cg + c_2 \gc ]\fpm ,
\end{equation}
where $c_1$ and $c_2$ are constants.  (That $c_1$ and $c_2$ are not more
generally independent functions of $\th_1$ and $\th_2$ is easily seen;
the constraints \ref{eVii} and \ref{eViii} do not allow it.)
$c_1$ and $c_2$ can be fixed easily.  Suppose
$|\alpha\rangle = (1/2,1/2,1/2,1/2)$ and
$|\beta\rangle = (1/2,-1/2,-1/2,-1/2)$.  Then by rewriting the Majorana
fields as Dirac fields, one readily finds that
\begin{equation}\label{eVxliii}
_{\pm}f^1_{\alpha\beta} =  i _{\pm}f^2_{\alpha\beta} .
\end{equation}
This in turn forces $c_2 =0$.  We can then set $c_1 = 1$ as we 
are uninterested in an overall normalization.

Lorentz invariance demands
\begin{equation}\label{eVxliv}
\fpm = e^{\pm(\th_1+\th_2)/4}\fo (\th_{12}) .
\end{equation}
$\fo$ is then constrained through \ref{eVii},
\begin{eqnarray}\label{eVxlv}
\fo (-\th ) &=& -S_2(i\pi - \th )
{2\pi i + 3\th \over 5\pi i - 3\th} \fo (\th ) \cr\cr
&=& -S_2 (\th ) \fo (\th ) ,\end{eqnarray}
where $S_2$ is given in \ref{eIVxxii}, and through \ref{eViii},
\begin{equation}\label{eVxlvi}
\mp i ({_{\pm}R_{\beta a}})\fo (\th - 2\pi i) = \fo (-\th ) ,
\end{equation}
where $_{\pm}R_{\alpha a} = \pm i$ is the braiding phase.
Given that $S_2(\th)$ can be written as
\begin{eqnarray}\label{eVxlvii}\cr
S_2 (\th ) &=& - \exp \bigg[ \int^\infty_0 {dx \over x}
s({x\th \over i\pi}) G_f(x) \bigg] ,\cr\cr\cr
G_f(x) &=& {e^{-7x/6} + 2c(x/6) \over c(x/2)},
\end{eqnarray}
we see that $\fo$ takes the form
\begin{equation}\label{eVxlviii}
\fo (\th ) =  \exp \bigg[ \int^\infty_0 {dx \over x}
{G_f(x) \over s(x)} \sin^2({x\over 2\pi}(i\pi + \th))\bigg] .
\end{equation}
As two kinks of opposite chirality form a fermionic bound
state,  $\fkk$ should have a pole at
$\th_2 = \th_1 + i u^a_{\alpha\beta} = \th_1 + i2\pi/3$.
Thus $\fkk$ becomes,
\begin{eqnarray}\label{eVxlix}\cr
\fkkt &=& A_F \cg {e^{\pm (\th_1 + \th_2)/4} \over c(\th_{12} ) + 1/2}
\times\cr\cr\cr
&& \hskip -.5in \exp \bigg[ \int^\infty_0 {dx \over x}
{G_f(x) \over s(x)} \sin^2({x\over 2\pi}(i\pi + \th_{12}))\bigg] ,
\end{eqnarray}
where $A_F$ is some normalization with mass dimension $[m]^{1/2}$.

To determine the phase of $A_F$ we again employ a hermiticity
condition:
\begin{equation}\label{eVl}
\fkkt^* = {_\pm f}^a_{\bar{\beta} \bar{\alpha}}(\th_2 - i\pi , 
\th_1 - i\pi).\end{equation}
This implies the phase of $A_F$ is $e^{\pm i\pi/4}$.  Scaling this phase out
leaves us with the final form for $\fkk$:
\begin{eqnarray}\label{eVli}\cr
\fkkt &=& A_F e^{\pm i\pi/4}
\cg {e^{\pm (\th_1 + \th_2)/4} \over c(\th_{12} ) + 1/2}\times\cr\cr\cr
&& \hskip -.5in \exp \bigg[ \int^\infty_0 {dx \over x}
{G_f(x) \over s(x)} \sin^2({x\over 2\pi}(i\pi + \th_{12}))\bigg] .
\end{eqnarray}

\subsection{Two Particle Form Factors for Kink Operators}

\newcommand{\fk}{\psi^\alpha_\pm}
\newcommand{\kfk}{{_\pm f}^\alpha_{a\beta}}
\newcommand{\kfkt}{{_\pm f}^\alpha_{a\beta}(\th_1,\th_2)}
\newcommand{\kkf}{{_\pm f}^\alpha_{\beta a}}
\newcommand{\kkft}{{_\pm f}^\alpha_{\beta a}(\th_1,\th_2)}

At the two particle level, a right/left kink field, $\fk$, will
couple to a kink $A_\beta (\th )$ of chirality opposite to $\alpha$
together with
a Gross-Neveu fermion.  So we consider form factors given by
\begin{equation}\label{eVlii}
\kfkt = \langle \fk (0) A_\beta (\th_2 ) A_a (\th_1 ) \rangle .
\end{equation}
For $\kfk$ to be invariant under \son , it must take the form
\begin{equation}\label{eVliii}
\kfkt = \cg \fpm + \gc {\tilde f}_\pm (\th_1,\th_2) .
\end{equation}
We can set ${\tilde f}_\pm (\th_1,\th_2)$ to 0.
The constraints upon the form
factors \ref{eVii} and \ref{eViii} lead to different functional forms for
$\fpm$ and ${\tilde f}_\pm (\th_1,\th_2)$.  By triality we expect
these forms to match that of $\fkk$ and only $\fpm$ does.

We must also consider $\kkf$, the form factor obtained from $\kfk$ by
reversing the order of particles.  By triality both $\kfk$ and $\kkf$ are
expressible in terms of form factors of type $\fkk$.  Because $\fkk$
is symmetric in $\alpha ,\beta$, we must then have
\begin{equation}\label{eVliv}
\kfk = \kkf ,
\end{equation}
and so $\kkft = (C\gamma^a)_{\alpha\beta} \fpm $.

Lorentz invariance demands that
\begin{equation}\label{eVlv}
\fpm = e^{\pm (\th_1 +\th_2)/4} \fo (\th_{12}) .
\end{equation}
This ansatz is again constrained by \ref{eVii} and \ref{eViii} in conjunction
with the kink-fermion S-matrix \ref{eIVxx}:
\begin{eqnarray}\label{eVlvi}\cr
\fo (\th - 2\pi i ) &=& -S_2(\th ) \fo (\th ) ;\cr\cr
\fo (\th - 2\pi i ) &=& \fo (-\th ) .\end{eqnarray}
That these equations match \ref{eVxliv} and \ref{eVxlv} is a
reflection of triality.  Given the bound structure of $\kfk$
is the same as $\fkk$, we can write down
\begin{eqnarray}\label{eVlvii}
\kfk (\th_1,\th_2) &=& \kkf (\th_1,\th_2) \cr\cr
&=& -A_F e^{\pm i\pi/4}\cg {e^{\pm (\th_1 + \th_2)/4} 
\over c(\th_{12}) + 1/2}\cr\cr\cr
&& \hskip -.6in \times \exp \bigg[ \int^\infty_0 {dx \over x}
{G_f(x) \over s(x)} \sin^2({x\over 2\pi}(i\pi + \th_{12})) \bigg] ,
\end{eqnarray}
where the normalization relative to $\fkk$ has been fixed using triality.

\subsection{One Particle Form Factor for Kink Fields}

\newcommand{\kk}{{_\pm}f^\alpha_\beta}
\newcommand{\kkt}{{_\pm}f^\alpha_\beta (\th )}
\newcommand{\Cab}{C_{\alpha\beta}}

The kink field, $\fk$, will couple to the anti-kink 
$C_{\alpha\beta}A_\beta$.
Hence we consider the form factor,
\begin{equation}\label{eVlviii}
\kkt = \langle \fk (0) A_\beta (\th )\rangle ,
\end{equation}
which must take the form
\begin{equation}\label{eVlix}
\kkt = c_\pm e^{\pm \th /2} \Cab .
\end{equation}
To determine the constant, we use the two particle form
factor $\kfk$.  Let $|\alpha\rangle = (1/2,1/2,1/2,1/2) $ and
$|\beta\rangle = (-1/2,-1/2,-1/2,-1/2)$.  $A_\beta$ can be written
as
\begin{equation}\label{eVlx}
i g^\beta_{a\gamma} A_\beta (\th) = {\rm res}_{\delta = 0}
A_a(\th + \delta + i \bar{u}^a_{\gamma\bar{\beta}})
A_\gamma(\th - i \bar{u}^{\bar{\gamma}}_{a\bar{\beta}}),
\end{equation}
where $a=1$, $|\gamma\rangle = (1/2,-1/2,-1/2,-1/2)$.  Again the $u$'s
mark out poles in S-matrix indicative of bound states.  Here they are given
by
\begin{equation}\label{eVlxi}
\bar{u}^a_{\gamma\bar{\beta}} = \bar{u}^{\bar{\gamma}}_{a\bar{\beta}}
= \pi - u^{\bar{\gamma}}_{a\bar{\beta}} = \pi/3 .
\end{equation}
$g^\beta_{a\gamma}$ can be determined up to a phase
from the fermion-kink S-matrix
as before to be
\begin{equation}\label{eVlxii}
g^\beta_{a\gamma} = {1\over 2}
\bigg(\sqrt{3\pi} {\Gamma (5/3) \over \Gamma (7/6)}\bigg)^{1/2} .
\end{equation}
Given that,
\begin{equation}\label{eVlxiii}
i g^\beta_{a\gamma} \kkt = {\rm res}_{\delta = 0} ~ {_\pm}
f^\alpha_{\gamma a} (\th - i\pi/3 , \th + \delta + i\pi/3) ,
\end{equation}
together with the hermiticity constraint,
${_\pm}f^{\bar{\alpha}}_\alpha(\th )^* =  (\mp i){_\pm}
f^\alpha_{\bar{\alpha}}(\th )$,
we find (up to a sign)
\begin{eqnarray}\label{eVlxiv}\cr
c_\pm &=& {4 \over \sqrt{3}} A_F e^{\pm i\pi/4}
\bigg( \sqrt{3\pi} {\Gamma (5/3) \over \Gamma (7/6)}\bigg)^{-1/2}\cr\cr\cr
&& \hskip .2in \times
\exp \bigg[ - \int^\infty_0 {dx \over x}{G_f(x) \over s(x)} 
s^2(x/6)\bigg] .
\end{eqnarray}

\subsection{One Particle Form Factor for the Gross-Neveu Fermions}

\newcommand{\fff}{{_\pm}f^a_b}

At the one particle level, the fermion field, $\ff$ will couple only
to $A_a (\th )$.  Hence we have
\begin{equation}\label{eVlxv}
\fff = \langle\ff (0) A_b(\th ) \rangle = d_\pm e^{\pm \th/2} \delta_{ab} .
\end{equation}
To determine the constant $d_\pm$, we use triality.  Fixing $a=2$, it is
easy to show under the triality transformation \ref{eIIxxv},
\begin{equation}\label{eVlxvi}
{_\pm}f^a_a = {1\over 2} ( {_\pm}f^\alpha_{\bar{\alpha}}
 + {_\pm}f^{\bar{\alpha}}_\alpha),
\end{equation}
where $\alpha = (1/2,1/2,1/2,1/2)$ and so $d_\pm = c_\pm$.

\subsection{Summary}

Here we summarize the results of this section for quick reference:

\vskip .2in
\noindent{\bf Two Particle Form Factors:}
\vskip .2in
For the $SO(8)$ currents, $\cop$, we have
\begin{eqnarray}\label{eVlxvii}
\fcft &\equiv& \langle G_\mu^{ab} (0) A_b (\th_2) A_a (\th_1) \rangle\cr\cr
&=& i A_G (\delta_{ac}\delta_{bd} - \delta_{ad}\delta_{bc})
f_\mu (\th_1,\th_2) ,\cr\cr
\fckt &\equiv& \langle G_\mu^{ab} (0) A_\beta (\th_2) 
A_\alpha (\th_1) \rangle\cr\cr
&=& i {A_G \over 2} \cs f_\mu (\th_1,\th_2) ,
\end{eqnarray}
where
\begin{eqnarray}\nonumber
f_{\mu}(\th_1,\th_2) &=&
(e^{(\th_1+\th_2)/2} - (-1)^\mu e^{-(\th_1+\th_2)/2})\cr\cr
&& \times{s(\th_{12}/2) \over c(\th_{12}) - 1/2}\cr\cr
&& \times \exp \bigg[\int^\infty_0 {dx \over x} {G_c(x) \over s(x)}
\sin^2({x\over 2\pi}(i\pi + \th_{12}))\bigg] ,\cr\cr\cr
G_c(x) &=& 2 {c(x/6) - s(x/6)e^{-2x/3} \over c(x/2)}.
\end{eqnarray}
\begin{equation}\label{eVlxviii}\end{equation}

For the Gross-Neveu fermions, $\ff$, and the kinks, $\fk$, we have
\begin{eqnarray}\nonumber
\fkkt &=& 
\langle \psi^a_\pm (0) A_\beta (\th_2) A_\alpha (\th_1)\rangle\cr\cr
&=& A_F e^{\pm i\pi/4} \cg \fpm ,\cr\cr
\kfkt &=& \kkft \equiv \langle \psi^\alpha_\pm (0) A_\beta (\th_2) 
A_a (\th_1)
\rangle \cr\cr
&=& -A_F e^{\pm i\pi/4} \cg \fpm ,
\end{eqnarray}
\begin{equation}\label{eVlxix}\end{equation}
where
\begin{eqnarray}\label{eVlxx}
\fpm &=& {e^{\pm (\th_1 + \th_2)/4} \over c(\th_{12} ) + 1/2}\cr\cr
&& \hskip -.5in \times \exp \bigg[\int^\infty_0 {dx \over x}
{G_f(x) \over s(x)} \sin^2({x\over 2\pi}(i\pi + \th_{12}))\bigg] ,\cr\cr
G_f(x) &=& {2 c(x/6) + e^{-7x/6} \over c(x/2)}.
\end{eqnarray}

\vskip .2in
\noindent{\bf One Particle Form Factors:}
\vskip .2in

For the currents, $\cop$, we have
\begin{eqnarray}\label{eVlxxi}
_\mu f^{ab}_{\{cd\}} (\th ) &\equiv&
\langle G^{ab} (0) A_{\{cd\}} (\th ) \rangle\cr\cr
&=& i A_G
(\delta_{ac}\delta_{bd} - \delta_{ad}\delta_{bc})(e^\th - 
(-1)^\mu e^{-\th})\cr\cr
&&\hskip -.2in  \times {1 \over \sqrt{3}}
\bigg( 2\sqrt{3\pi} {\Gamma (2/3) \over \Gamma (1/6)}\bigg)^{-1/2}\cr\cr
&&\hskip -.2in 
\times \exp \bigg[-\int^\infty_0 {dx \over x} {G_c(x) \over s(x)} 
s^2(x/3)\bigg] ,
\end{eqnarray}
and for the fermions, $\ff$, and kinks, $\fk$,
\begin{eqnarray}\label{eVlxxii}\cr
\fff &\equiv& \langle \psi^a_\pm (0) A_b (\th )\rangle =
c_\pm e^{\pm \th/2} \delta_{ab} ,\cr\cr
\kk &\equiv& \langle \psi^\alpha_\pm (0) A_\alpha ( \th ) \rangle =
c_\pm e^{\pm \th/2} C_{\alpha\beta} ,
\end{eqnarray}
where the constant $c_\pm$ is
\begin{eqnarray}\label{eVlxxiii}
c_\pm &=& {4\over \sqrt{3}} e^{\pm i\pi/4}
A_F \bigg(\sqrt{3\pi} {\Gamma (5/3) \over 
\Gamma (7/6)}\bigg)^{-1/2}\cr\cr\cr
&& ~ \times 
\exp \bigg[ - \int^\infty_0 {dx \over x}{G_f(x) \over s(x)} s^2(x/6)\bigg].
\end{eqnarray}

\section{Acknowledgments}
We would like to thank M.P.A. Fisher, L. Balents, H. Lin, N. MacKay, 
A. LeClair,
D. Duffy, D. Scalapino, G. Sierra, and I. Affleck for useful discussions.
This work has been supported by
NSERC of Canada and the NSF through the Waterman Award under
grant number DMR-9528578 (R.K.) and by the A.P. Sloan Foundation (A.W.W.L.).

\end{multicols}

\end{document}